%% file: main.tex
\documentclass[%
  reprint,
superscriptaddress,
showpacs,preprintnumbers,
nofootinbib,
 amsmath,amssymb, 
 aps,
 prd%,
floatfix,
]{revtex4-2}
\usepackage{amsmath,amstext,amssymb}
\usepackage[T1]{fontenc}
\usepackage{apjfonts} 
\usepackage{xcolor}
\usepackage{soul}
\usepackage{comment}
\usepackage{graphicx}
\usepackage[normalem]{ulem}
\usepackage[export]{adjustbox}
\usepackage[colorlinks=true,allcolors=blue]{hyperref}
\usepackage{orcidlink}
\usepackage{multirow}
\usepackage{lineno}
\usepackage[toc,page]{appendix}
\graphicspath{{figures/}}

\newcommand{\mnras}{MNRAS}

\newcommand{\aap}{Astronomy \& Astrophysics}

\begin{document}

\title{Kilonova emission from GW230529 and mass gap neutron star-black hole mergers}

\author{K.~Kunnumkai~\orcidlink{0009-0000-4830-1484}}
\author{A.~Palmese~\orcidlink{0000-0002-6011-0530}}%\email{palmese@cmu.edu}
\affiliation{McWilliams Center for Cosmology and Astrophysics, Department of Physics, Carnegie Mellon University, Pittsburgh, PA 15213, USA}
\author{M.~Bulla~\orcidlink{0000-0002-8255-5127}}
\affiliation{Department of Physics and Earth Science, University of Ferrara, via Saragat 1, I-44122 Ferrara, Italy}
\affiliation{INFN, Sezione di Ferrara, via Saragat 1, I-44122 Ferrara, Italy}
\affiliation{INAF, Osservatorio Astronomico d’Abruzzo, via Mentore Maggini snc, 64100 Teramo, Italy}
\author{T. Dietrich~\orcidlink{0000-0003-2374-307X}}
\affiliation{Institut f\"{u}r Physik und Astronomie, Universit\"{a}t Potsdam, Haus 28, Karl-Liebknecht-Str. 24/25, 14476, Potsdam, Germany}
\affiliation{Max Planck Institute for Gravitational Physics (Albert Einstein Institute), Am M\"{u}hlenberg 1, Potsdam 14476, Germany}
\author{A. M. Farah~\orcidlink{0000-0002-6121-0285}}
\affiliation{Department of Physics, University of Chicago, Chicago, IL 60637, USA}
\author{P. T. H. Pang~\orcidlink{0000-0001-7041-3239}}
\affiliation{Institute for Gravitational and Subatomic Physics (GRASP), Utrecht University, Princetonplein 1, 3584 CC Utrecht, The Netherlands}
\affiliation{Nikhef, Science Park 105, 1098 XG Amsterdam, The Netherlands}

%~ 225 words for abstract PRD letter with tot 4,500 words for rest of paper

%%Letter justification: The GW230529 event was recently announced by the LIGO/Virgo/KAGRA collaboration. It is important and timely for the gravitational wave follow up community searching for kilonova signatures to know what to expect from this kind of events now, as it is possible other similar events will be detected during the ongoing O5 observing run. 

\begin{abstract}

The detection of the gravitational wave event GW230529, presumably a neutron star-black hole (NSBH) merger, by the LIGO-Virgo-KAGRA (LVK) Collaboration marks an exciting discovery for multimessenger astronomy. The black hole (BH) has a high probability of falling within the ``mass gap'' (mg) between the neutron star (NS) and the BH mass distributions. Because of the relatively low primary mass, this system has a higher likelihood of producing an electromagnetic counterpart than previously detected NSBH mergers. We analyze the potential kilonova (KN) emission from GW230529 and find that, if the source was an NSBH merger, there is a $\sim 2$–$28\%$ probability (depending on the assumed equation of state) that it produced a KN peaking at $\sim 1$ day post-merger with $g \lesssim 23.5$ and $i < 23$. Hence, it could have been detected by ground-based telescopes. If instead the event was a binary neutron star (BNS) merger, the probability of KN production drops to $\sim 0$–$10\%$. Motivated by these results, we simulate a broader population of mgNSBH mergers expected during the fifth LIGO/Virgo/KAGRA observing run (O5) and find a $2$–$3\%$ chance of KN production per event. Such KNe would typically be fainter than GW230529, with $g \lesssim 26$ and $i \lesssim 25$. Based on these findings, DECam-like instruments may be able to detect up to $\sim 70\%$ of future mgNSBH KNe, corresponding to $1-2$ multimessenger mgNSBH per year in O5.

\end{abstract}


\maketitle

\section{Introduction}

In May 2023, the LIGO-Virgo-KAGRA (LVK) Collaboration began the fourth gravitational wave (GW) observing run (O4) with the Advanced LIGO~\cite{LIGOScientific:2014pky}, Advanced Virgo~\cite{VIRGO:2014yos}, and KAGRA~\cite{KAGRA:2020tym} detector network. On 29 May 2023, the network detected an exceptional signal: GW230529 \cite{230529_LVK}. This event is of particular interest because it is likely a neutron star-black hole (NSBH) merger with a primary compact object of mass within $2.5-4.5~M_\odot$. This implies that the mass of the primary falls within the so-called lower ``mass gap'' (e.g. \cite{Bailyn_1998,Özel_2010,Farr_2011,Belczynski_2012}) between the anticipated mass ranges of the neutron stars (NSs) and that of the black holes (BHs). 

If GW230529 was indeed an NSBH merger, the neutron star may have been disrupted outside the black hole’s innermost stable circular orbit (ISCO), possibly producing an electromagnetic (EM) counterpart such as a short $\gamma$-ray burst (GRB) or a kilonova (KN), see \cite{Kyutoku:2021icp,Nakar:2019fza,Foucart:2021pau} and references therein. No GRB signatures were detected for this event \cite{230529_GRB}, possibly because the merger was observed off-axis, as suggested by GW constraints on the inclination angle. However, in contrast to GRBs, KNe, which are the ultraviolet-to-infrared counterparts of GWs powered by radioactive decay of the neutron-rich nuclei \cite{metzger_kilonovae}, are expected to be more isotropic and therefore could still be detectable even for off-axis mergers. 

Unfortunately, GW230529 was detected by only one detector, and its sky localization in low-latency with BAYESTAR~\cite{Singer:2015ema} covered about 24,200 square degrees at the 90\% credible interval (CI). Hence, no online KN search could be carried out successfully for this event \cite{Ahumada_2024}. However, archival searches for KNe using large-sky surveys may still provide constraints \cite{Pillas2025}. Furthermore, GW230529-like mergers are expected to occur at rates comparable to NSBH mergers with more asymmetric mass components \cite{230529_LVK}, making it interesting to explore the KN properties and detectability of such systems. A candidate GW event (S250206dm; \cite{2025GCNLVK,2025GCNLVK2,2025GCNLVK1,2025GCNLVK3}) with high probability of being an NSBH and possibly bearing a mass-gap black hole was indeed detected following GW230529, although KN searches did not reveal any promising electromagnetic counterpart \cite{Hu2025,Ahumada2025}.

In this work, we examine two questions: (1) what KN signatures could GW230529 have produced and (2) what KN properties should we expect from a realistic population of NSBH mergers with BHs in the lower mass gap. Section~\ref{section:Methodology} describes our methods for simulating KNe from GW230529 and the broader mass gap NSBH (mgNSBH) population. Section~\ref{section:Results} presents the results, along with comparisons to related studies \cite{zhu2024,230529_everything,matur2024}. Finally, Section~\ref{sec:conclusion} summarizes our conclusions.

\section{Methods}\label{section:Methodology}

\subsection{Kilonova simulations for GW230529}

Using the measured masses and observationally motivated priors on the mass distributions of black holes and neutron stars, the LVK Collaboration concluded that GW230529 most likely originated from the coalescence of a BH and a NS. However, the non-detection of tidal effects in the GW signal prevents a definitive classification of the source. To remain agnostic, we consider different possibilities for the binary’s nature and analyze each with the corresponding waveform-informed posteriors. Throughout, we use the GW230529 posterior samples released in Ref.~\cite{GW230529_posterior_samples_zenodo}.

\subsubsection{Posterior sets and waveform models}
We first assume GW230529 as an NSBH source and analyze two posterior sets:

(i) a binary black hole (BBH) waveform posterior, \texttt{Combined\_PHM\_highSpin} \cite{230529_LVK,LVKopendata}, constructed by combining the frequency-domain \texttt{IMRPhenomXPHM} \cite{PhysRevD.102.064001,PhysRevD.102.064002} and time-domain \texttt{SEOBNRv5PHM} \cite{PhysRevD.108.124036,PhysRevD.108.124035} models, with spin magnitudes restricted to $\chi_1,\chi_2 < 0.99$. This posterior does not include tidal disruption effects.

(ii) an NSBH waveform posterior, \texttt{IMRPhenomNSBH} \cite{PhysRevC.58.1804}, which incorporates tidal effects from the NS companion \cite{PhysRevD.100.044003} and assumes aligned spins with $\chi_1 < 0.5$ and $\chi_2 < 0.05$ for the primary and secondary, respectively. This model targets NSBH binaries with mass ratios $q=m_1/m_2 \sim 1$–$15$, where $m_{1}$ and $m_{2}$ denote the masses of the primary and secondary components. 

As an alternative hypothesis in which both components are neutron stars, we employ the posterior assuming the BNS waveform model \texttt{IMRPhenomPv2\_NRTidalv2} \cite{PhysRevD.100.044003}, which includes tidal effects from both bodies and uses spin restrictions $\chi_1 < 0.99$ and $\chi_2 < 0.05$. We do not report results for the low-spin prior case ($\chi_{1}, \chi_{2} < 0.05$), since those posteriors do not yield BNS KNe for any of the three EOSs considered here, as the primary mass samples lie above the maximum neutron star mass. 

Across all cases, we adopt the LVK default parameter-estimation priors without population-informed adjustments. We then compute the expected KN properties for each posterior sample set. For clarity, we use the following labels: under the NSBH assumption, the case where we assume the BBH waveform posterior \texttt{Combined\_PHM\_highSpin} is denoted ``NSBHs–BBHw'', while ``NSBHs–NSBHw'' refers to the case where we use the NSBH waveform posterior \texttt{IMRPhenomNSBH}. Under the BNS source assumption, the cases where we assume the BNS waveform posterior \texttt{IMRPhenomPv2\_NRTidalv2} are denoted ``BNSs–BNSw''. In this notation, “s” indicates the assumed source type (NSBH or BNS), and “w” indicates the waveform model.

\subsubsection{Ejecta prescriptions and kilonova mapping}

We employ the Nuclear Physics and MultiMessenger Astronomy (NMMA) framework \cite{Dietrich_2020,Pang_2023}, which connects GW source properties (e.g., compactness, mass ratio, and effective spin $\chi_{\rm eff}$, provided in the form of posterior samples) to KN ejecta properties using fitting formulae. The main ejecta components we model are the dynamical ejecta mass ($M_{\rm dyn}$) and the wind ejecta mass ($M_{\rm wind}$). 

For BNS mergers, we evaluate $M_{\rm dyn}^{\rm BNS}$ using the analytic prescription of \cite{Kr_ger_2020} and the disk mass $M_{\rm disk}^{\rm BNS}$ following \cite{Dietrich_2020}. In this case, $M_{\rm dyn}$ depends mainly on the mass ratio and compactness, while $M_{\rm wind}$ depends on the total mass and the threshold mass for prompt collapse.

For NSBH mergers, we use \cite{Kr_ger_2020} for $M_{\rm dyn}^{\rm NSBH}$ and \cite{Foucart_2018} for the remnant baryon mass $M_{\rm rem}^{\rm NSBH}$ (measured $\sim 10$\,s post-merger). Here, $M_{\rm dyn}$ and $M_{\rm rem}$ depend on the component masses, BH spin, NS baryonic mass, and NS compactness. We compute the disk mass as $M_{\rm disk}^{\rm NSBH}=M_{\rm rem}^{\rm NSBH}-M_{\rm dyn}^{\rm NSBH}$ and draw a Gaussian fraction of $M_{\rm disk}$ to obtain $M_{\rm wind}$ (see Appendix~\ref{sec:appendix} and Sec.~II.B of \cite{kunnumkai} for a more detailed discussion of the equations used and the ejecta parameters). Following \cite{Barbieri_2019,2024MNRAS.52711053M} and motivated by the numerical-relativity results of \cite{Foucart2019} for low-mass NSBH mergers, we cap the dynamical ejecta to $<50\%$ of the remnant mass.

We count a merger as KN-producing only if the total ejecta exceeds a fixed threshold: $M_{\rm ej,tot}\ge 10^{-4}\,M_\odot$ for both BNS and NSBH. Events below these limits are excluded from the KN-producing subset. The fraction of KN-producing mergers that are detectable is evaluated in Sec.~\ref{sec:kndetection}.

For our fiducial analysis, we adopt the maximum-posterior equation of state (EOS) from \cite{Huth:2021bsp} (Zenodo: \cite{huthzenodo}). Each EOS in the ensemble carries a probability weight (listed in \texttt{posterior\_probability.txt} in \cite{huthzenodo}) informed by astrophysical and nuclear-physics constraints. For every EOS, we compute the tidal deformability of a $1.4~M_\odot$ neutron star $\Lambda_{1.4}$ by interpolating its mass-tidal deformability relation, then form the \emph{weighted} distribution of $\Lambda_{1.4}$ using the provided posterior probabilities. The EOSs at the lower and upper $2.5\%$ quantiles of this weighted distribution define our “softer” and “stiffer” EOSs (corresponding to files \texttt{14836.dat} and \texttt{12592.dat} in \cite{huthzenodo} respectively), to show a set of physically motivated extremes while the median ($50\%$ quantile, corresponding to file \texttt{14512.dat} in \cite{huthzenodo}) defines the “fiducial” EOS.

We compute synthetic KN light curves with \texttt{POSSIS}–based models. For NSBH posteriors (NSBHs–BBHw and NSBHs–NSBHw), we use an extended Bu2019nsbh \cite{Anand_2020} grid, where we have added low-mass ejecta models down to $10^{-4}~M_\odot$, which are needed for the NSBHs in question. For BNS posteriors (BNSs–BNSw), we use an extended Bu2019lm \cite{Dietrich_2020} grid. Both models are built on \texttt{POSSIS} \cite{Bulla_2019,Bulla_2023}, a three-dimensional, time-dependent radiative-transfer code. The underlying \texttt{POSSIS} training grids assume radioactive $r$-process heating, wavelength and composition dependent opacities (lanthanide-poor wind vs.\ lanthanide rich dynamical ejecta), an axisymmetric geometry (quasi spherical wind plus an equatorial, lanthanide rich wedge), and anisotropic emission set by the viewing angle $\theta_v$. Within NMMA, surrogate models trained on these grids are able to return emulated light curve for a variety of input parameters. The NMMA surrogate has been retrained for this work using the extended POSSIS grids. In Bu2019nsbh, the surrogate requires 3 key inputs: $\log_{10}(M_{\rm dyn})$, $\log_{10}(M_{\rm wind})$, and the viewing angle $\theta_v$. In Bu2019lm, the default inputs are $\log_{10}(M_{\rm dyn})$, $\log_{10}(M_{\rm wind})$, $\theta_v$, and the half-opening angle $\phi$ of the lanthanide-rich dynamical ejecta. We generate light curves in the $ugrizy$ bands. 

\subsection{Kilonova simulations for mass gap NSBH mergers}

We compare the GW230529 posterior samples with a mock population of mass gap GW events generated using LVK O5 sensitivities from \cite{Abbott_2020} and \url{https://dcc.ligo.org/LIGO-T2000012-v2/public}: \texttt{AplusDesign} for the two LIGO detectors, \texttt{avirgo\_O5low\_NEW} for Virgo, and \texttt{kagra\_128Mpc} for KAGRA. We draw component masses and spins from the \textsc{Power Law + Dip + Break} (PDB) model \citep{fishbach_does_2020,Farah_2022,abbott_population_2023}. The PDB model enforces a mass ``dip'' between $\sim 2.2\text{–}6.0$ M$_{\odot}$, which is allowed to vary in depth and mass range. This dip between the expected NS and BH population will include events like GW230529 since it is not necessarily an empty mass gap. %Because the inference for the PDB model does not change significantly with the inclusion of GW230529 \cite{230529_LVK} we use the posteriors computed based on the third GW Transient Catalog (GWTC-3; \cite{gwtc3}). 
The employed pairing function ensures similar components' masses within a given binary. After sampling masses and spins from the PDB model, we follow \cite{Petrov_2022} and use a network signal-to-noise ratio of 8, with the criteria that at least one of the detectors detects the signal above the threshold; cf.~\cite{kunnumkai} for details of the simulation pipeline.

For this study, we focus on the detected NSBH subpopulation whose primaries lie in the mass gap, $\sim 2.2\text{--}6.0\,M_\odot$, and we simulate KN light curves for these events. We consider all NSBHs with black hole mass up to $6.0\,M_\odot$. We set spin-magnitude $\chi$ ranges based on component mass, following \cite{abbott_population_2023}:for components with masses $<2.5\,M_\odot$ we draw $\chi$ from $[0,\,0.4]$, while for components with masses $>2.5\,M_\odot$ we draw $\chi$ from $[0,\,1]$. The direction distribution of the spins is isotropic.

Since our compact object population (which is drawn based on the PDB model hyper-parameters posterior from GWTC-3) does not depend on a binary classification, we define appropriate mass ranges for NS/BH boundaries using Eq.~(12) of \cite{Breu_2016}, which gives the maximum mass of a uniformly rotating NS, in terms of the maximum mass of a non-rotating NS, Tolman–Oppenheimer–Volkoff (TOV) limit $M_{\rm TOV}$ \cite{PhysRev.55.374,Kalogera_1996}. This boundary is EOS dependent. For our fiducial EOS we adopt $M_{\rm TOV,fid} = 2.436~M_{\odot}$. For the softer and stiffer EOSs, we take $M_{\rm TOV,l95} = 2.069~M_{\odot}$ and $M_{\rm TOV,u95} = 2.641~M_{\odot}$, respectively.

\section{Results and Discussion}\label{section:Results}

\subsection{Kilonova production from GW230529}\label{section:resultsproduction230529}

\begin{table}
    \centering
    \begin{tabular}{c|c|c|c|c}
        \hline
        \hline
        Event/ & Source type & \multicolumn{3}{c}{\% producing KN} \\
        \cline{3-5}
        population & & softer EOS & fiducial EOS & stiffer EOS \\
        \hline
        \hline
        \multirow{3}{*}{GW230529} 
            & BNSs-BNSw    & 1.1 & 6.2 & 11 \\
            & NSBHs-BBHw   & 1.7 & 13 & 20 \\
            & NSBHs-NSBHw  & 3.9 & 22 & 28 \\
        \hline
        LVK O5 & mgNSBH      & 2.1 & 2.8 & 3.3 \\
        \hline
        \hline
    \end{tabular}
    \caption{Percentage of GW230529 posterior samples (by posterior model) or simulated O5 mass gap NSBH events expected to produce KN emission for each EOS considered. The BNSs–BNSw row corresponds to the scenario in which GW230529 was a BNS; NSBHs–BBHw and NSBHs–NSBHw are related to the NSBH scenario under the assumption of different waveform models.}
    \label{tab:kn_percentage}
\end{table}

\begin{figure*}[htpb!]
    \centering
    \includegraphics[scale=0.175]{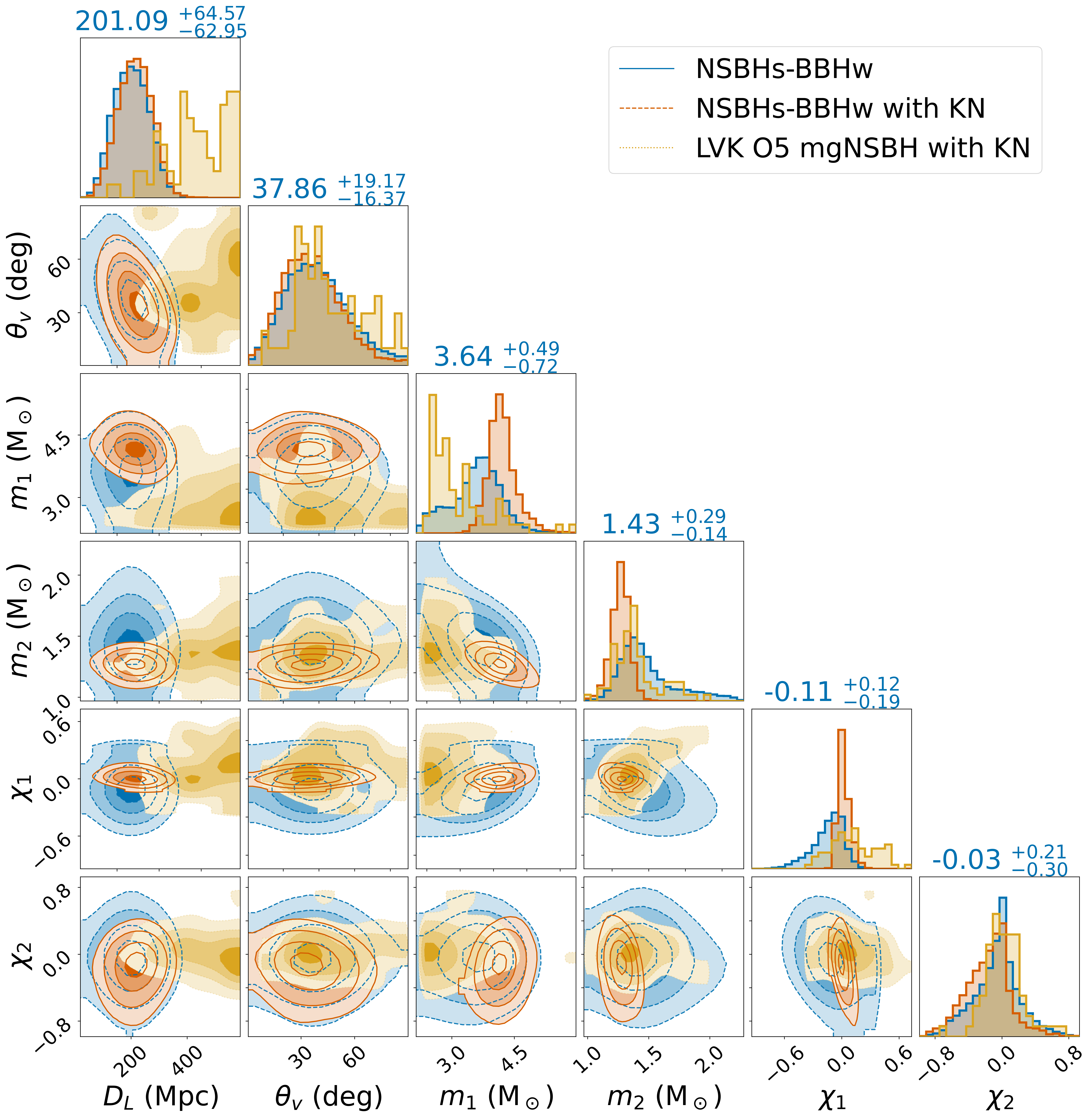}
    \includegraphics[scale=0.175]{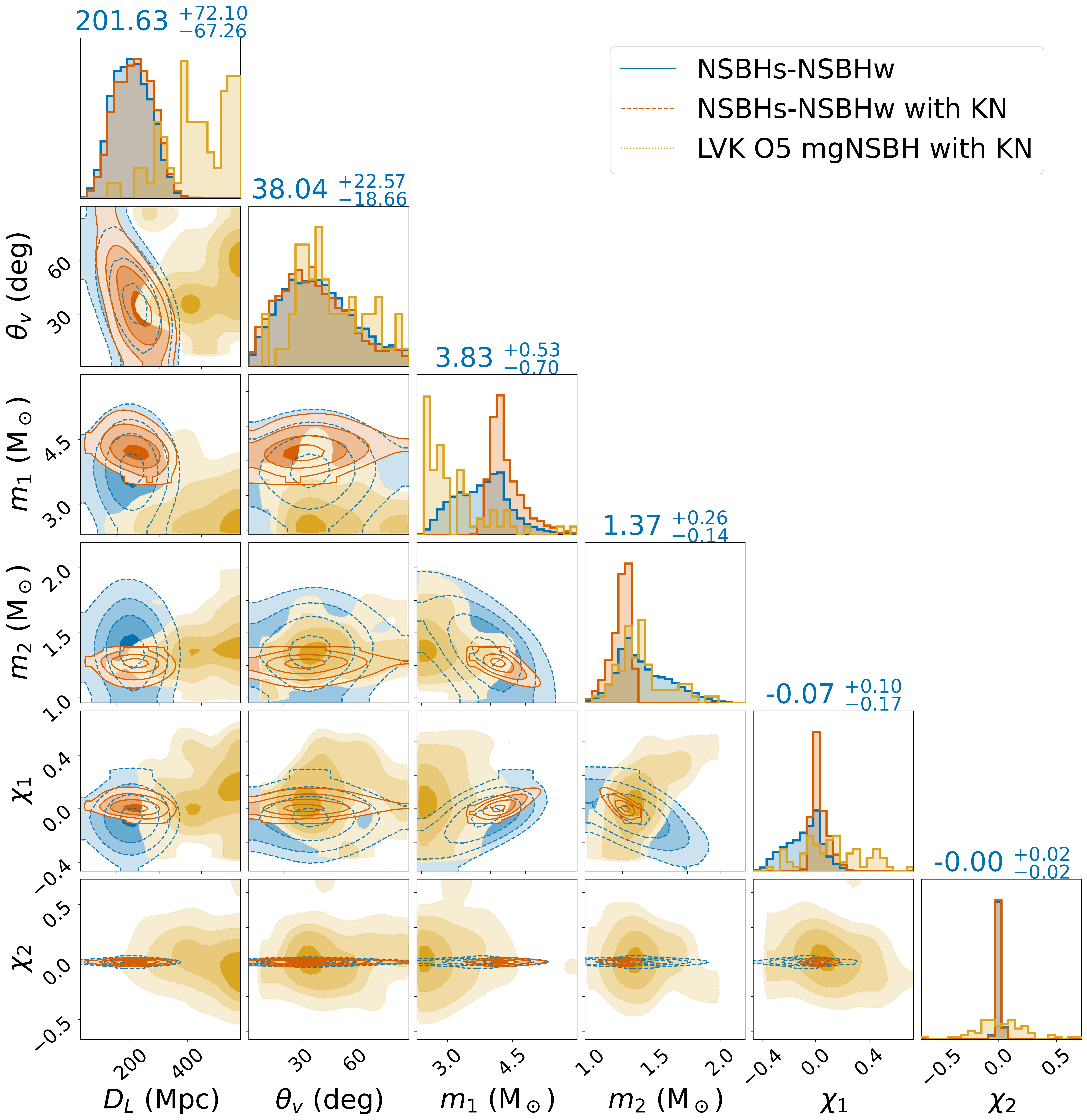}
    \caption{Posterior distributions of GW230529 binary parameters under the NSBH assumption and adopting the BBH waveform (NSBHs–BBHw; left) and the NSBH waveform (NSBHs–NSBHw; right) models. Blue shows the full posterior; red shows the subset of samples that produce a kilonova assuming the fiducial EOS. Yellow indicates the simulated LVK O5 mass-gap NSBH population that produces a KN at distances 0-600~Mpc, shown for comparison with GW230529. The contours indicate the 1,2,3$\sigma$ of the distributions as well as the entire distribution including any outliers, from darker to lighter color. Note that the KN-producing samples of GW230529 (red) are shifted relative to the peak of the simulated O5 KN producing population (yellow). This is expected as the chirp mass driven $m_1$–$m_2$ degeneracy in GW230529 pushes KN production into the high-$m_1$, low-$m_2$ region, whereas the simulated O5 population, lacking this degeneracy and favoring more equal mass systems, peaks at smaller $m_1$ and larger $m_2$.}
    \label{fig:post_kn_models_BBH_NSBH}
\end{figure*}

\begin{figure*}[htpb!]
    \centering
    \includegraphics[scale=0.175]{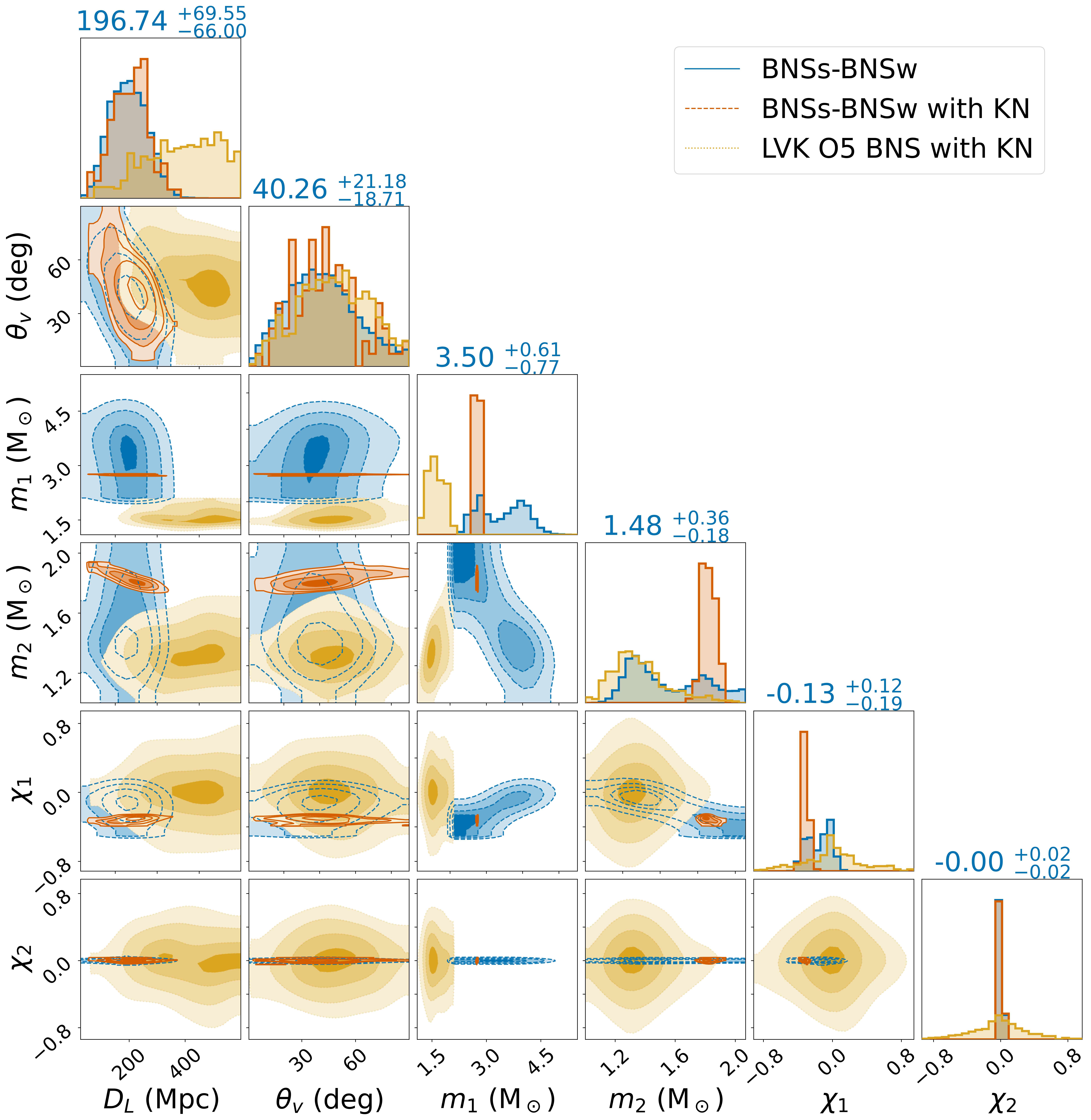}
    \includegraphics[scale=0.175]{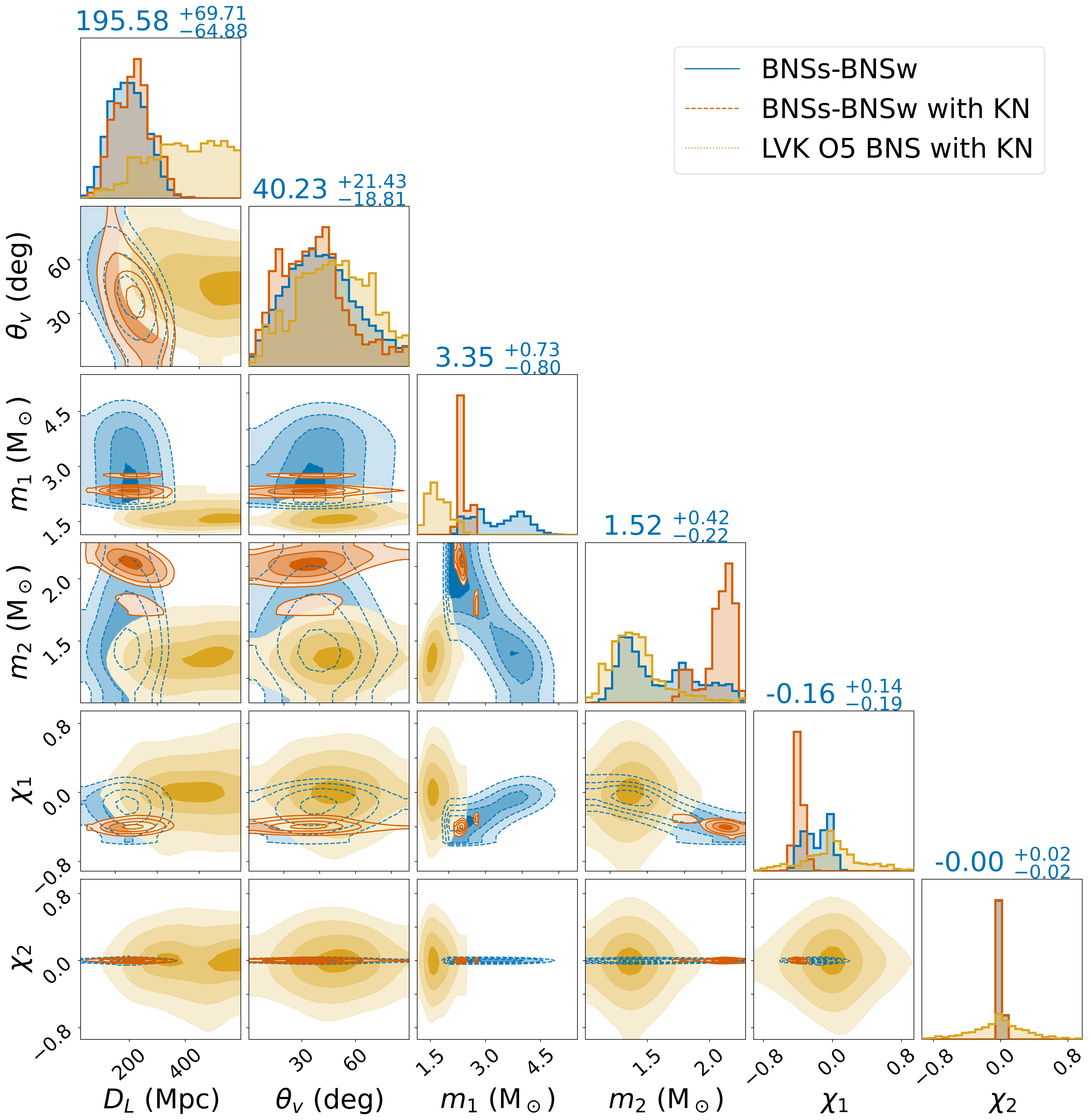}
    \caption{Posterior distributions of GW230529 binary parameters under the BNS assumption (BNSs–BNSw). Blue shows the full posterior; red shows the subset of samples that produce a KN for the softer EOS (left) and the fiducial EOS (right). Yellow indicates the simulated LVK O5 BNS population that produces a KN at luminosity distances of 0-600~Mpc, shown for comparison. The values above the histograms refer to the GW230529 binary posteriors. The contours indicate the 1,2,3$\sigma$ of the distributions as well as the entire distribution including any outliers, from darker to lighter color. Only the lowest mass, fastest spinning primaries in the GW230529 posterior are classified as neutron stars; due to degeneracies, these correspond to high mass ($\sim 2M\odot$) secondaries. By contrast, in the generic O5 BNS population, KN production is favored by lower-mass secondaries.}
    \label{fig:post_kn_models_BNS}
\end{figure*}

We estimate the probability that GW230529 produced a KN by analyzing, for each posterior set, the fraction of samples that yield non-negligible ejecta under different waveform models and equations of state (EOS). We also examine how KN production varies across the binary parameter space. Figure~\ref{fig:post_kn_models_BBH_NSBH} and Figure~\ref{fig:post_kn_models_BNS} show the posterior distributions of GW230529 for luminosity distance, viewing angle, component masses, and spins, adopting the \texttt{NSBHs-BBHw}, \texttt{NSBHs-NSBHw}, and \texttt{BNSs-BNSw} models. Blue contours denote the full posterior, while red contours mark the subset of samples that produce a KN. For the BNSs–BNSw case, we display both the softer and fiducial EOSs because the differences are more pronounced than in the NSBH case. Table~\ref{tab:kn_percentage} summarizes the fraction of KN-producing samples for each model and EOS.

Only a subdominant fraction of samples produces a KN in all scenarios. We attribute this to the degeneracy between $m_{1}$ and $m_{2}$, as a redshifted chirp mass is measured from the GW data, and individual component masses are not as well constrained. Posteriors that favor lower $m_2$ (which typically produces more ejecta mass than a more massive secondary, all other parameters being fixed) are bound to larger values of $m_1$, which disfavors tidal disruption and hence KN production. In addition, the effective spin is likely negative or close to zero, with $\sim 72\%$ probability that the primary spin is anti-aligned \cite{230529_LVK}. Anti-aligned spin increases the BH ISCO radius relative to a prograde BH of the same mass, further suppressing NS disruption and KN production in NSBH systems. As a result, only a specific subset of posterior samples can produce a KN. The size of this subset depends on how much posterior probability lies at low $m_2$, on the spin configuration of the NS and BH, and on the EOS. These factors can increase the maximum NS mass and the spin of the BH, which affects the ISCO radius; cf.~e.g.~\cite{Foucart:2020ats,Kyutoku:2021icp} and references therein.

As we progress from softer to stiffer EOSs, the range of $m_2$ that permits KN production shifts to higher masses, because stiffer EOSs allow more massive NSs (see Figure~\ref{fig:post_kn_models_BNS} for the BNSs–BNSw case). Due to the $m_1$-$m_2$ degeneracy, this shift drives $m_1$ toward lower values in the KN-producing subset. The same qualitative trend appears across all of the three cases considered (BNSs–BNSw, NSBHs–NSBHw, NSBHs–BBHw). 

\subsubsection{GW230529 as an NSBH}

For what concerns GW230529 as an NSBH, the NSBH waveform posterior (NSBHs–NSBHw) yields $\sim 4$–$28\%$ KN-producing samples, while the BBH waveform posterior (NSBHs–BBHw) yields $\sim 2$–$20\%$. In both cases, stiffer EOSs maximize the probability. The difference between waveform assumptions arises primarily from the spin priors, which affects both the maximum allowed NS mass and the BH ISCO radius. The BBHw posterior permits strongly negative BH spins: about $8.3\%$ of samples have $\chi_1<-0.4$, compared to $<0.5\%$ for the NSBHw posterior, which enlarges the ISCO radius and suppresses KN production. By contrast, the NSBHs-NSBHw posterior excludes such extreme retrograde spins and assigns greater weight to low secondary masses ($56\%$ of NSBHs-NSBHw samples versus $43\%$ for NSBHs-BBHw are at $m_2\lesssim 1.4\,M_\odot$), conditions that enhance KN production. Finally, as expected, the normalized luminosity distance and viewing angle distributions of the KN-producing samples are similar to those in the input posteriors, since these extrinsic parameters do not control whether a KN forms; we show them to facilitate comparison with the broader mgNSBH population.

\subsubsection{GW230529 as a BNS}

For the BNS posterior, the primary spin $\chi_1$ can take larger magnitudes than the secondary spin, which in principle permits more massive NSs to exist and produce a BNS KN provided the spin is anti-aligned, as the data do not favor the aligned case. Therefore, the EOS that allows the largest region of parameter space with fast, anti-aligned primary spin, given the $m_1$-$m_2$-$\chi_1$ degeneracy produces the highest KN fraction. As expected, the softer EOS yields the lowest KN production probability ($\sim 1\%$), which increases for the fiducial ($\sim 6\%$) and stiffer ($\sim 10\%$) EOSs.

Under the softer EOS, only a specific configuration of component masses and spins permits a KN: the primary is as massive as allowed by the spin prior while still remaining an NS, as illustrated in the left panel of Figure~\ref{fig:post_kn_models_BNS}. More negative primary spins or lower $m_1$ are disfavored because they would require an even more massive $m_2$, which is limited by the secondary’s spin prior $\chi_2<0.05$; consequently, $m_2$ saturates near $M_{\rm TOV,l95}$. For the fiducial EOS, the secondary can reach higher mass, which in turn allows the primary to assume more negative spins.

\subsubsection{Ejecta mass predictions}

\begin{figure*}
    \includegraphics[width=0.32\linewidth]
    {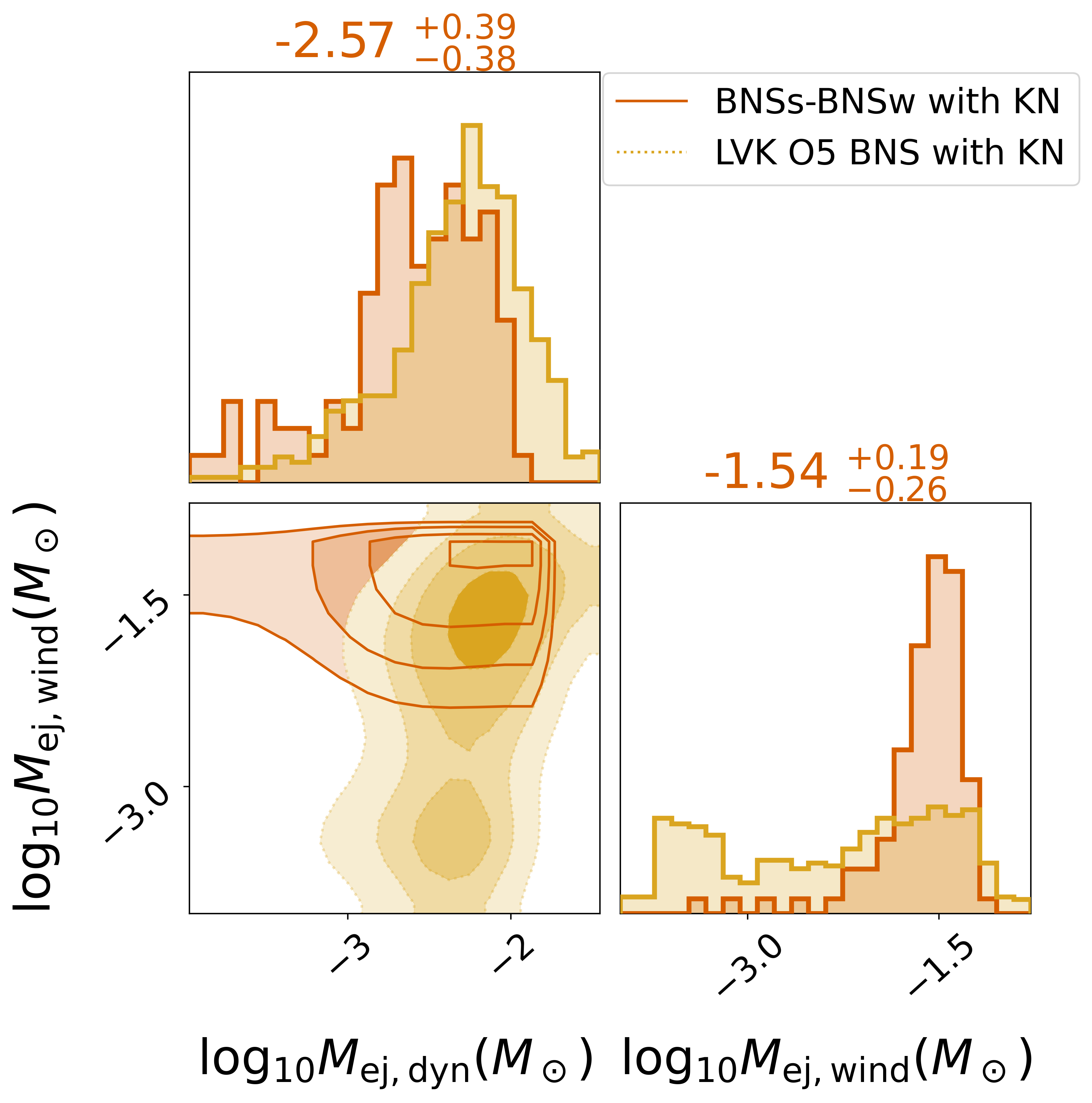}
    \includegraphics[width=0.32\linewidth]{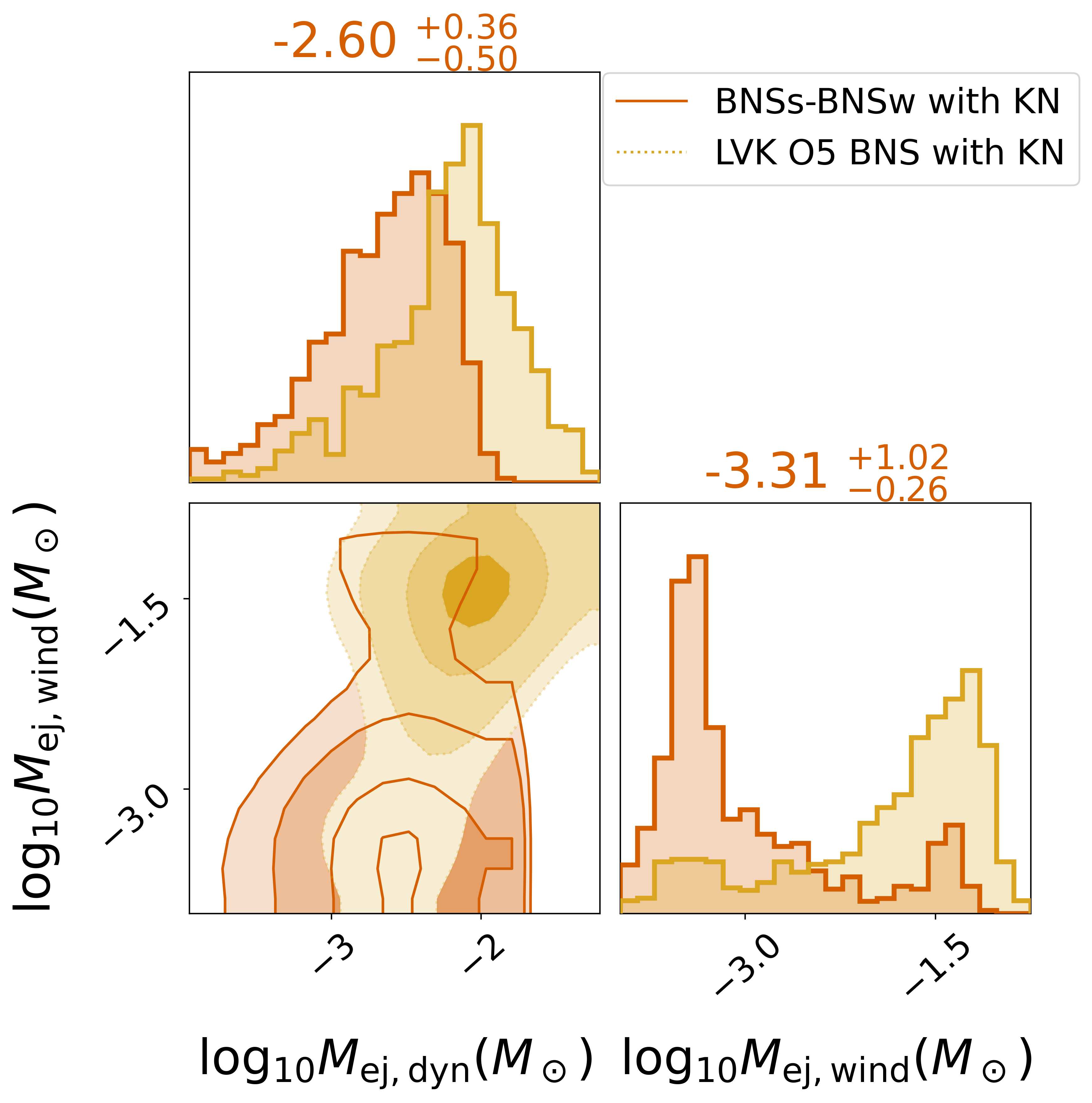}
    \includegraphics[width=0.32\linewidth]{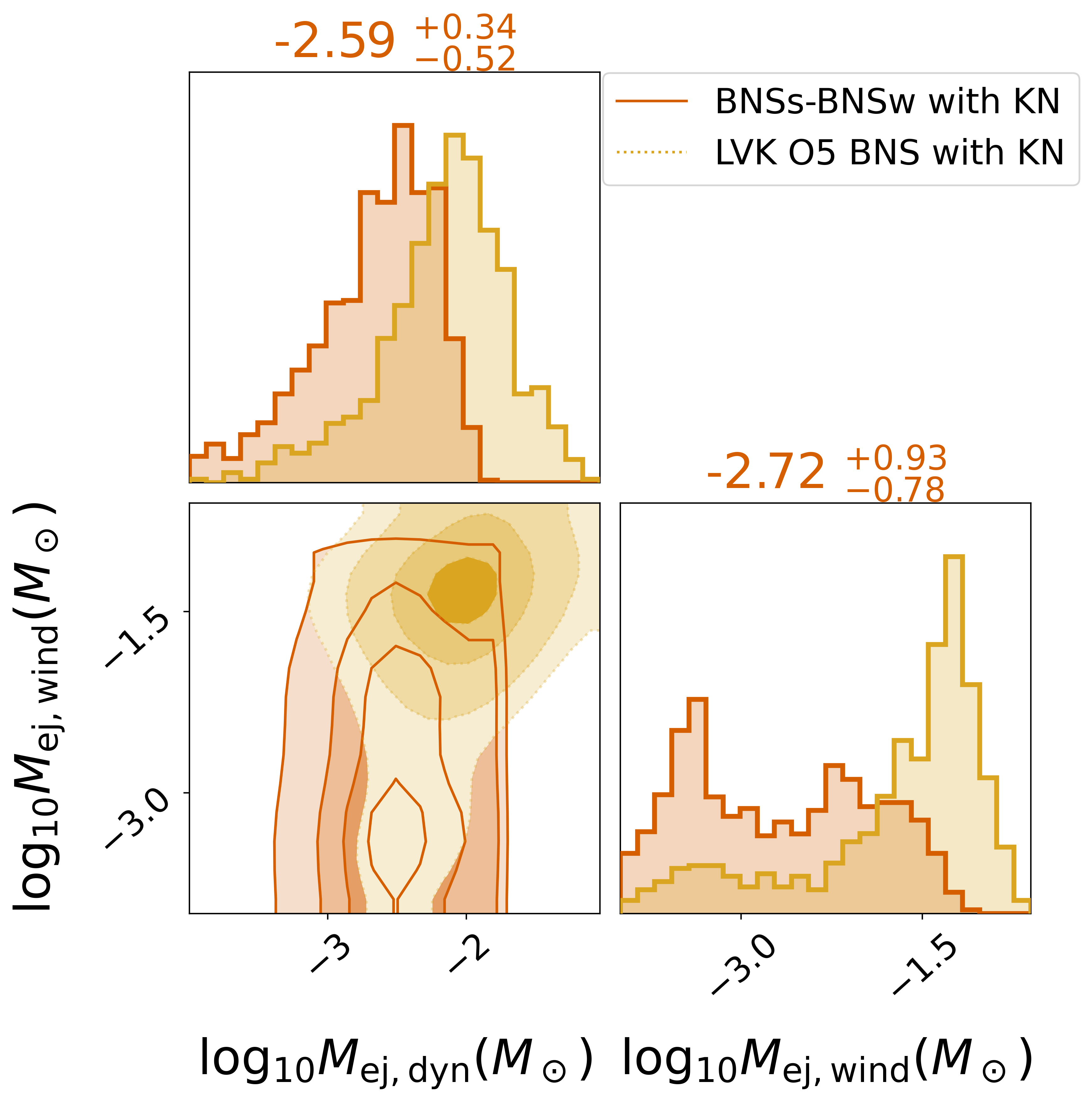}
    \caption{GW230529  cornerplots of dynamical and wind ejecta mass, in red, for the case of NSBHs-BBHw KN production assuming the GW230529 BBH posterior (top panels), and 
    BNS KN from the BNS posterior (bottom panels), a softer (left panels), fiducial (middle panels), and stiffer (right panels) EOS. The yellow contours show the expected population from mgNSBH (top) BNS mergers (bottom) expected to be detected at O5 sensitivity.  The contours indicate the 1,2,3$\sigma$ of the distributions as well as the entire distribution including any outliers, from darker to lighter color.}
    \label{fig:ejecta}
\end{figure*}

We now discuss the ejecta masses predicted for the KN-producing samples. For GW230529 as an NSBH, $M_{\rm wind}$ is in the range $(1-2)\times 10^{-3}\,M_\odot$ (median of the distribution) and smallest for the softer EOS and largest for the stiffer EOS, while the dynamical ejecta median spans $(4.0$–$7.2)\times 10^{-3}\,M_\odot$ for the NSBHw posterior and $(4.0$–$5.2)\times 10^{-3}\,M_\odot$ for the BBHw posterior (top panels of Figure~\ref{fig:ejecta}). These values agree with numerical–relativity estimates for low-mass NSBH binaries of $\sim 4\times10^{-3}$ to $5\times10^{-2}\,M_\odot$ \cite{Foucart2019}\footnote{Despite this general agreement, the BH spin orientation and component masses in GW230529 likely differs from those considered in \cite{Foucart2019}.}. 

\begin{figure*}
    \centering
    \includegraphics[width=0.33\linewidth]{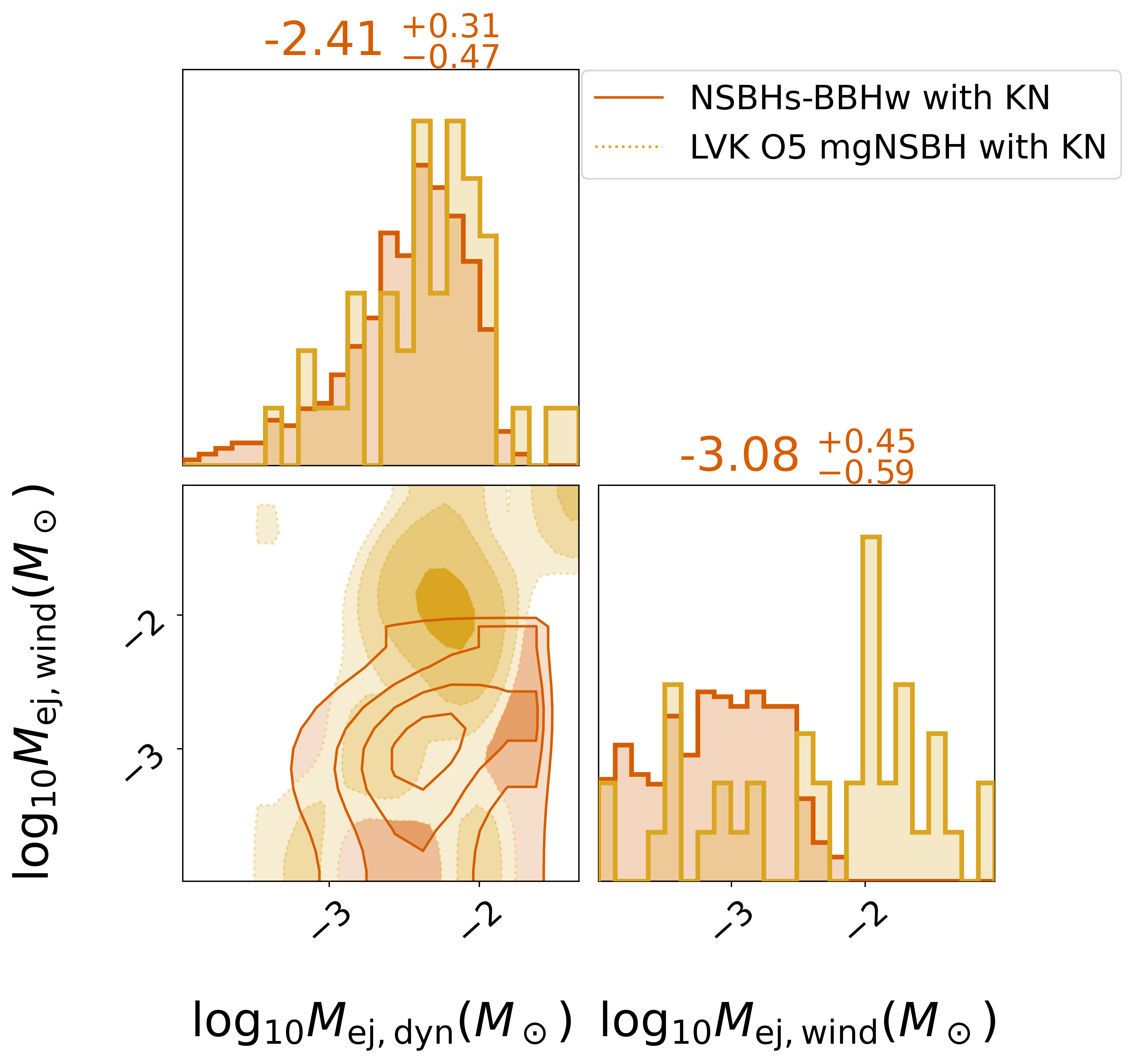}
    \includegraphics[width=0.33\linewidth]{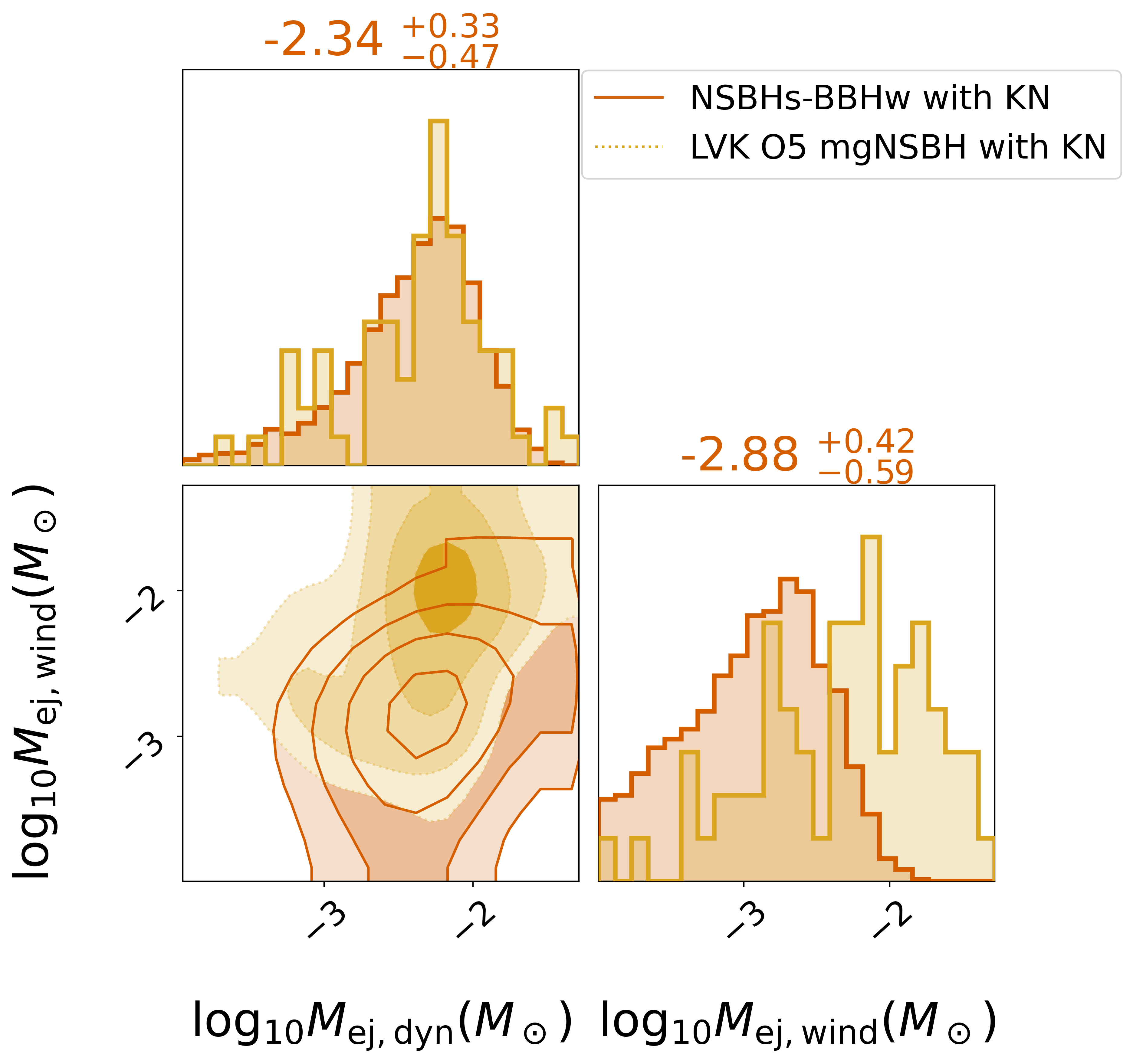}
    \includegraphics[width=0.33\linewidth]{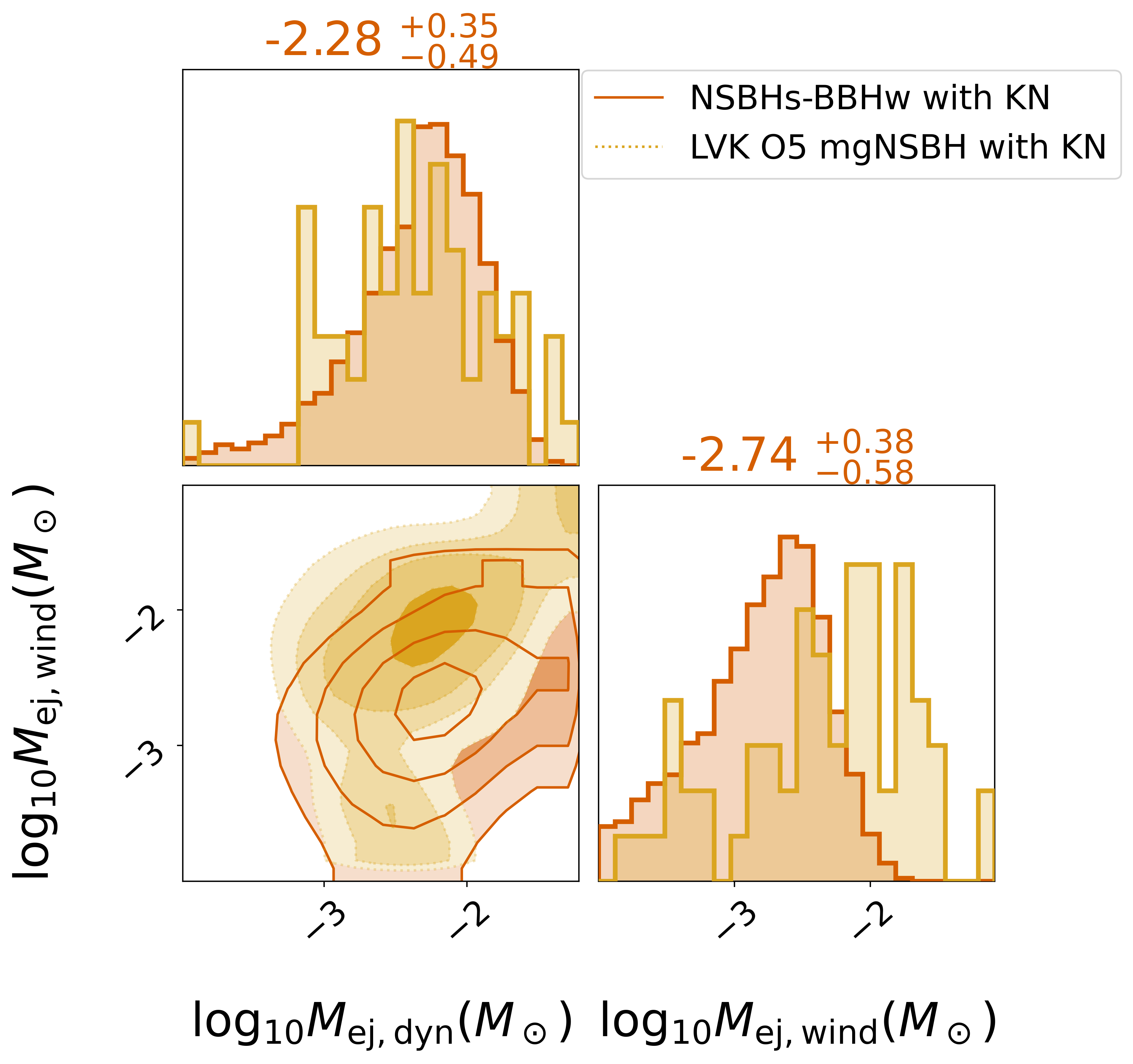}
\end{figure*}

For GW230529 as a BNS, the softer EOS is expected to yield more massive wind than dynamical ejecta, with wind ejecta masses of order $10^{-2}\,M_\odot$. In contrast, for the fiducial and stiffer EOSs, the BNS typically produces more dynamical than wind ejecta, with the former peaking near $3\times 10^{-3}\,M_\odot$ for all EOSs. This behavior reflects the BNS disk mass (from which $M_{\rm wind}$ is derived), which depends on the total mass and the prompt collapse threshold and is therefore highly EOS-sensitive. Although the fiducial and stiffer EOSs allow larger $M_{\rm tot}$ than the softer EOS, they do not produce disproportionately large disks. By comparison, the dynamical ejecta mass is most sensitive to the mass ratio; because the BNS KN-producing samples do not span a wide range in mass ratio, the dynamical ejecta remains similar across EOSs (see \cite{Dietrich_2020,Kr_ger_2020} for the governing relations). The dynamical ejecta distribution also resembles that of the NSBH case. Given the relatively large masses inferred for GW230529, its BNS dynamical ejecta is typically lower than for our O5 BNS population at similar distances  shifted by $\sim 0.3$–$1.5$\,dex toward smaller values (Figure~\ref{fig:ejecta}).

\subsection{Kilonova production from a mass gap NSBH merger population}

In this section, we predict KN production for LVK O5 mass-gap NSBH  mergers and compare those results with the potential KN emission from GW230529. We also include O5 BNS KN predictions at distances comparable to GW230529. Figure ~\ref{fig:post_kn_models_BBH_NSBH} and \ref{fig:post_kn_models_BNS} show the corresponding O5 populations (yellow contours and histograms).

\begin{figure*}[htpb!]
    \centering
    \includegraphics[scale=0.2685]{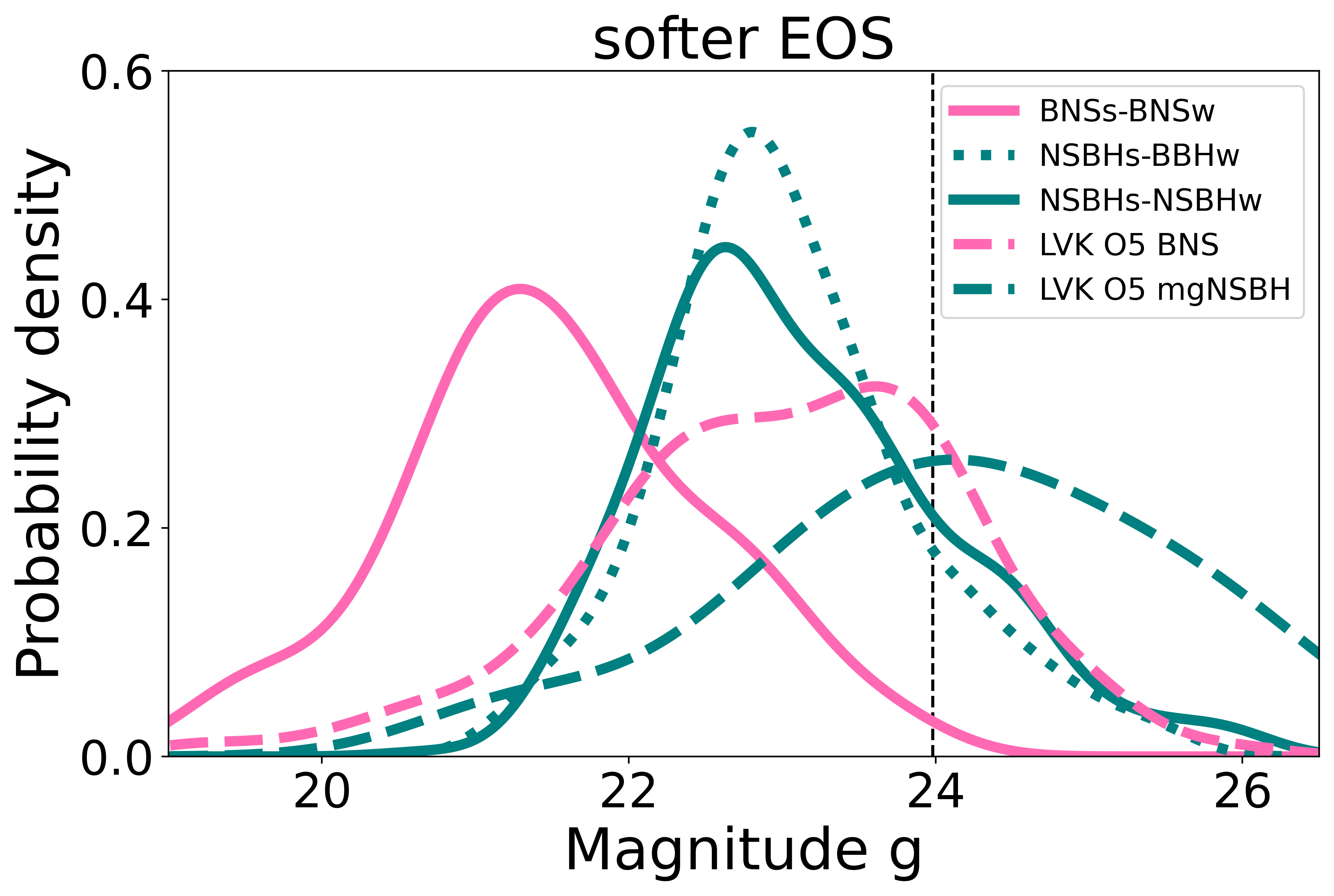}\includegraphics[scale=0.2685]{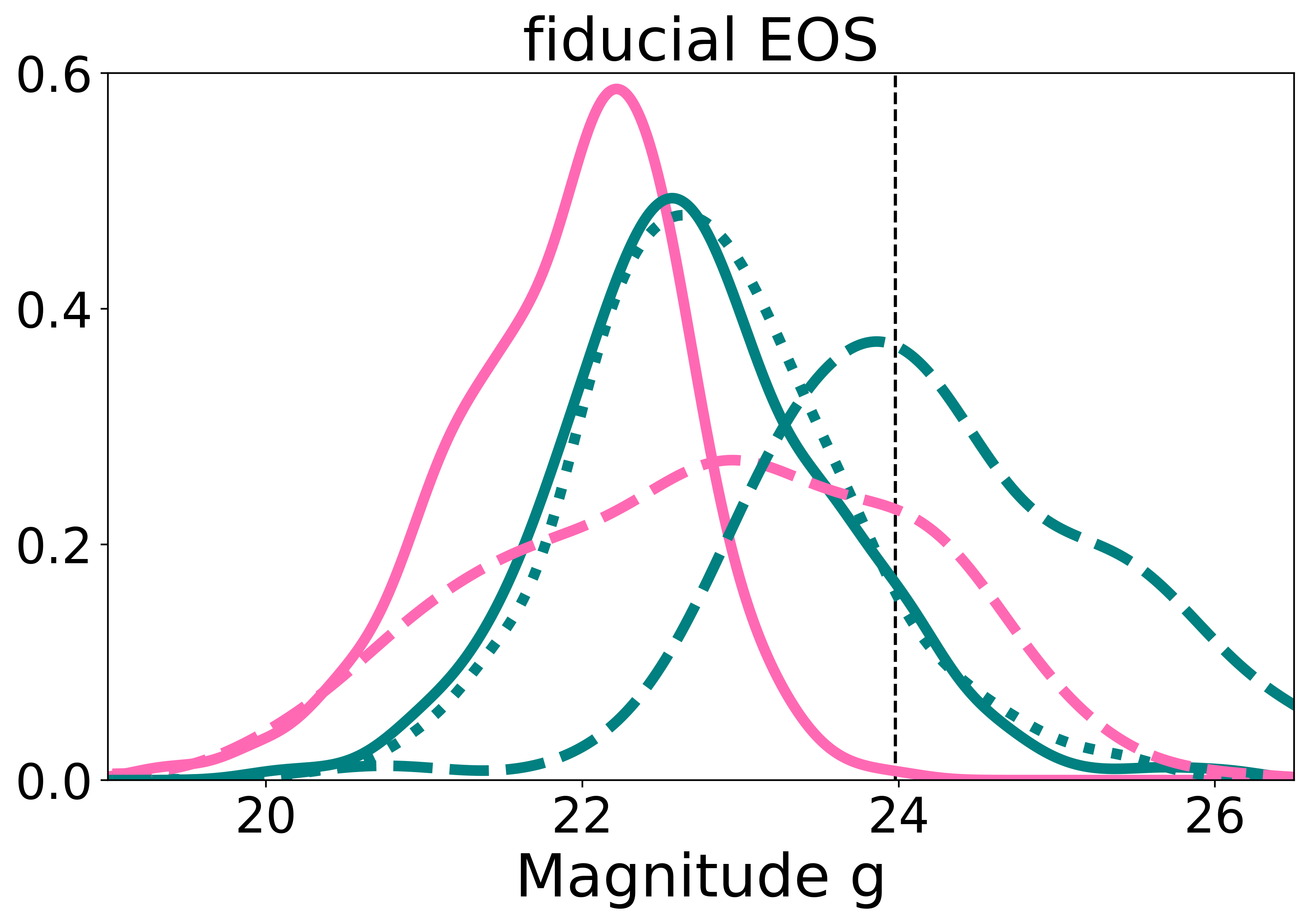}\includegraphics[scale=0.2685]{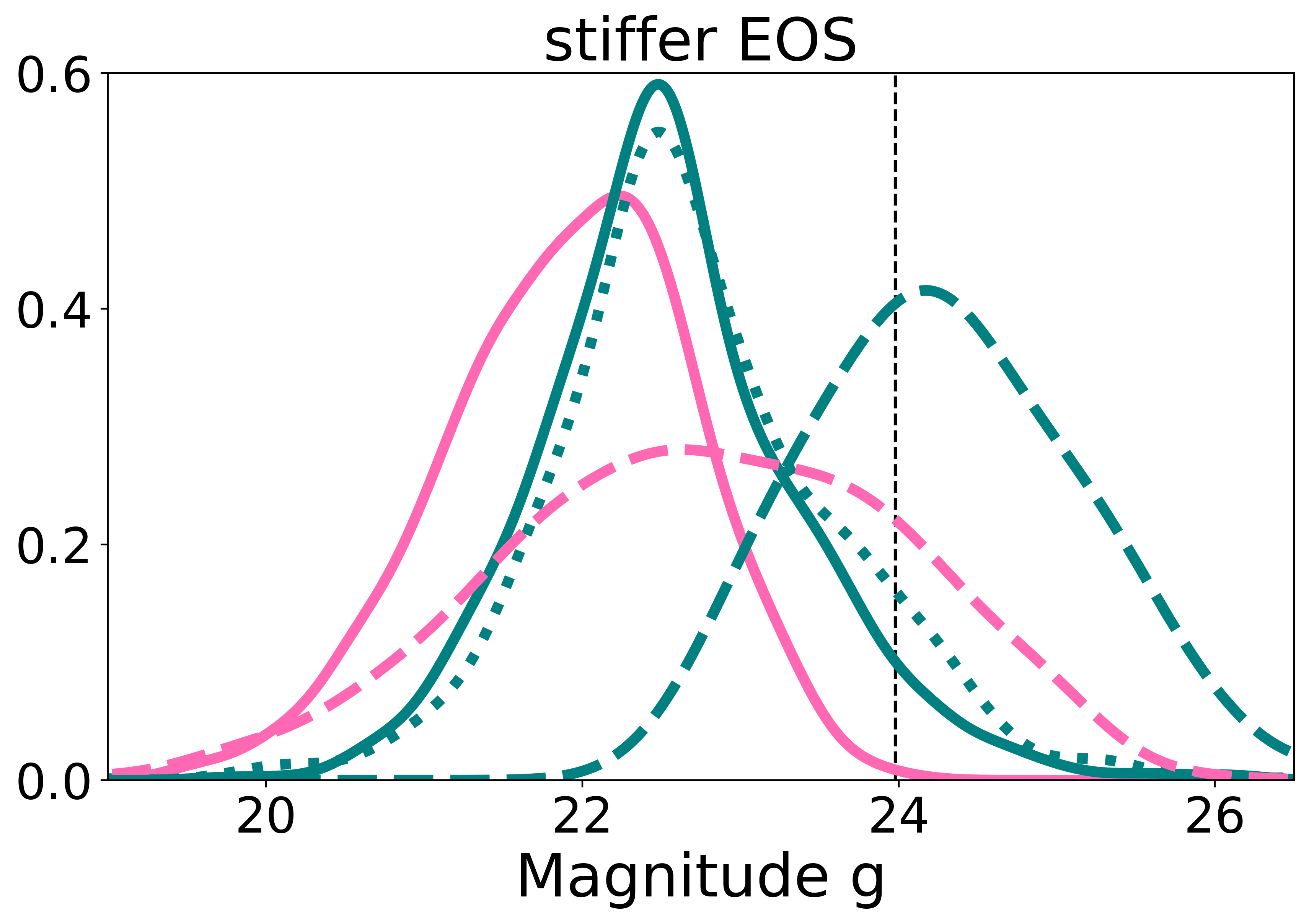}
    
    \includegraphics[scale=0.2685]{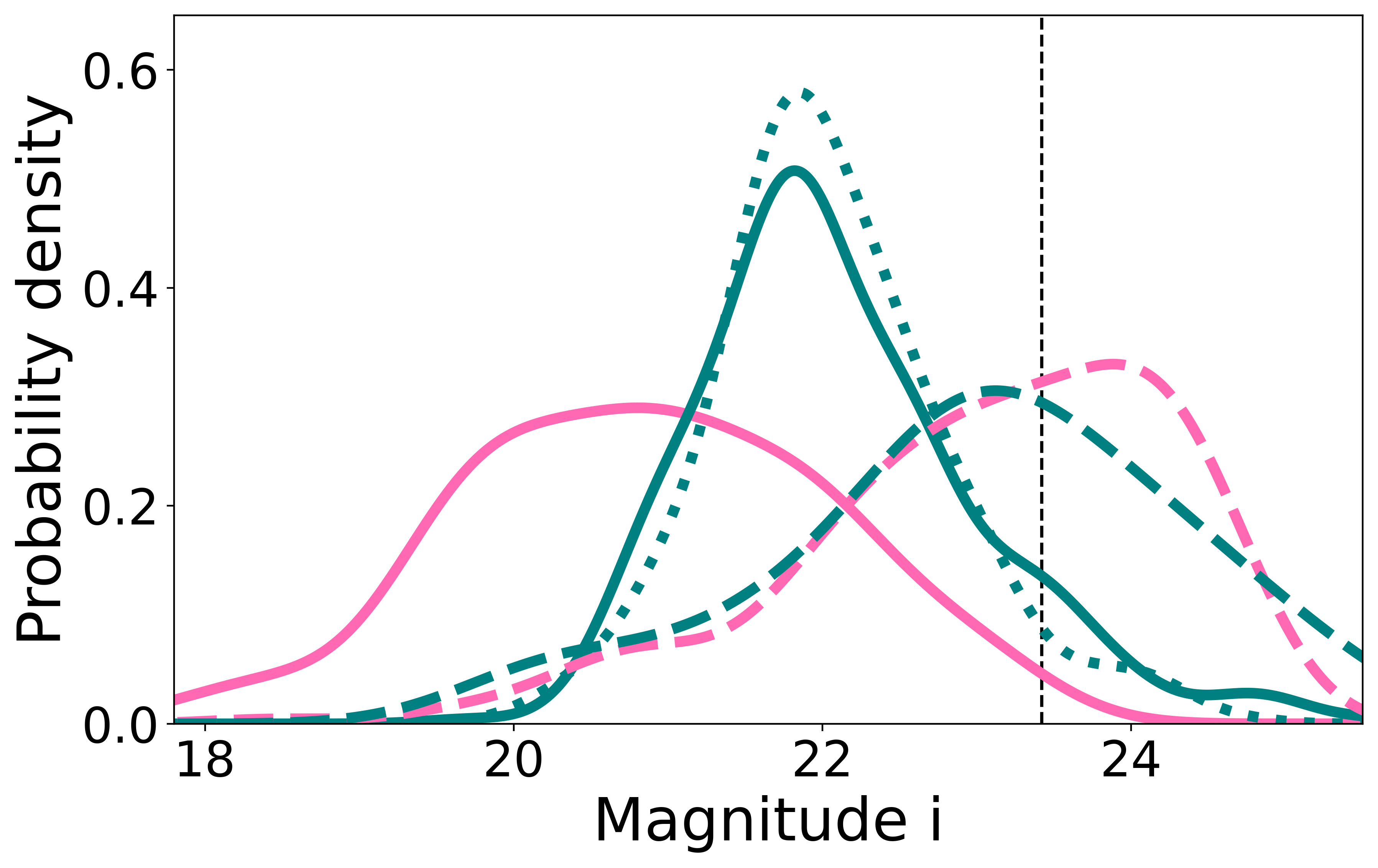}\includegraphics[scale=0.2685]{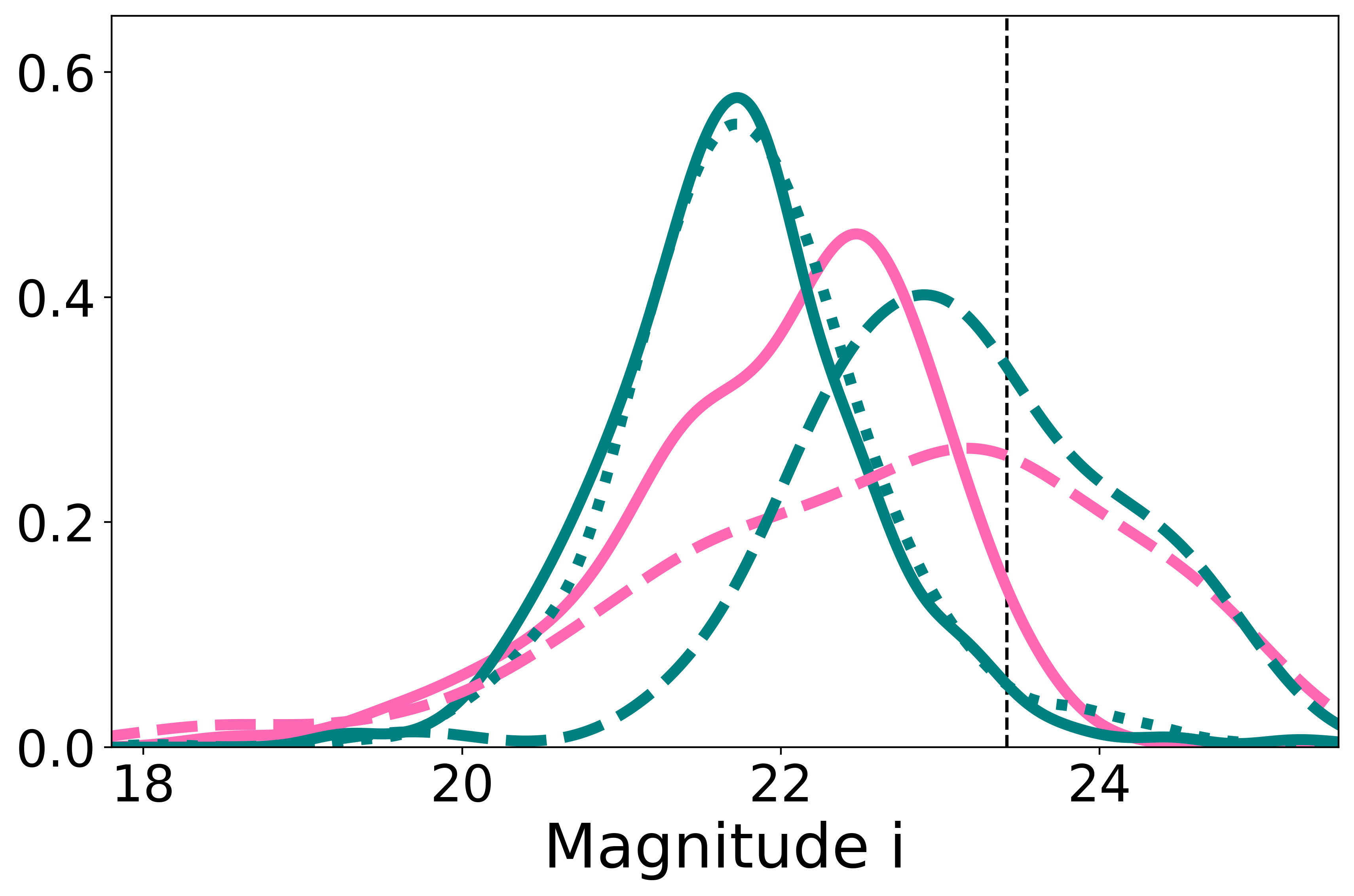}
    \includegraphics[scale=0.2685]{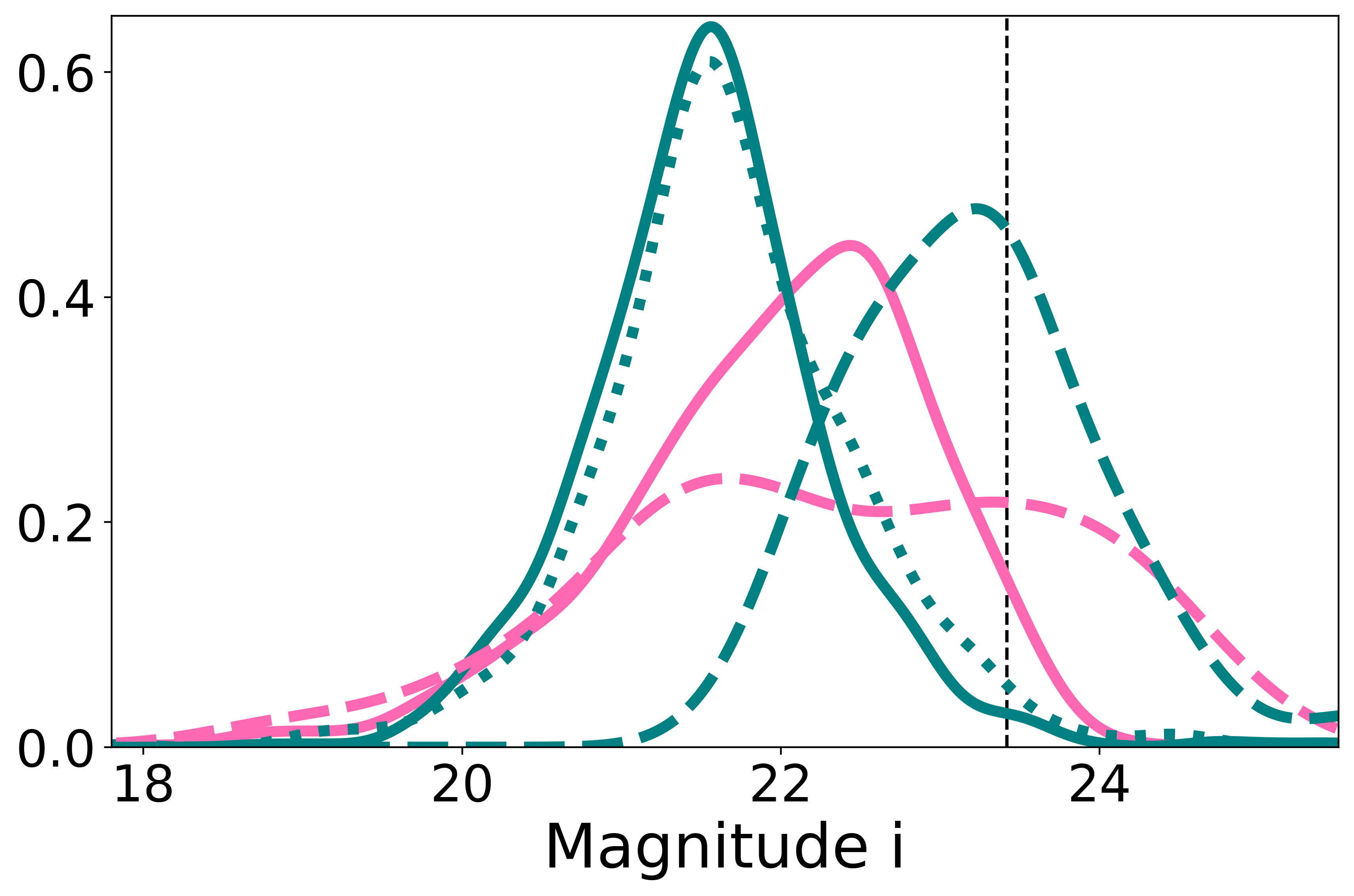}
    \caption{Simulated KN magnitudes in $g$ (top) and $i$ (bottom) at 1~day post-merger. Solid lines show GW230529 predictions for different posterior models; dashed lines show LVK O5 simulations for mgNSBH and BNS mergers. Left: softer EOS; middle: fiducial EOS; right: stiffer EOS. The black dotted line marks a reference DECam $5\sigma$ depth for a 90\,s exposure in each filter.}
    \label{fig:hist_day1}
\end{figure*}

We start by analyzing the KN emission from the simulated LVK O5 mass–gap events. Table~\ref{tab:kn_percentage} reports the percentage of NSBH mergers that produce a KN for each EOS. Similar to the GW230529 posteriors, the softer EOS yields the smallest KN fraction ($2\%$), while the fiducial and stiffer EOSs result in a KN production fraction of $\sim 3\%$. In Figure~\ref{fig:post_kn_models_BBH_NSBH}, the KN-producing subsets show a clear structure in spin and mass that mirrors the behavior seen for GW230529, but with some differences, which we describe in the following subsections.

\subsubsection{Spin distribution}

For the KN-producing samples under the GW230529 NSBH assumptions, $\chi_{1}$ shows a preference for an anti-aligned spin with respect to the binary angular momentum in both NSBHs–BBHw and NSBHs–NSBHw subsets, whereas the LVK O5 mgNSBH KN population prefers more aligned $\chi_1$ (about $88\%$ of our simulated events are aligned for the softer EOS and $74\%$ for the stiffer EOS). This is because the majority of the GW230529 posterior samples (83\% for BBHw and 72\% for NSBHw) do not allow for positive spin of the primary, whereas the O5 input distribution is uniform in spin magnitude and isotropic in orientation, which for events that produce a KN leads to a mild shift toward aligned $\chi_1$ in the O5 population.

For $\chi_2$, the O5 distribution spans a wide range with similar support for aligned and anti-aligned orientations, while the GW230529 NSBHs–BBHw KN-producing subset shows a preference for anti-aligned secondary spin. This preference reflects the three–way degeneracy between $\chi_1$, $m_1$, and $\chi_2$ along with the tendency for lower $m_1$ to favor disruption. We observe the same qualitative patterns across all three EOS choices.

\subsubsection{Mass distribution}

The $m_1$-$m_2$ degeneracy that shapes KN production for GW230529 favors lower $m_2$ and higher $m_1$ in the KN-producing subset. In contrast, the O5 mgNSBH population occupies a different region of parameter space. Because the adopted pairing function (based on GWTC–3) preferentially yields more equal-mass binaries and because merger rates within the dip are suppressed as higher masses are reached, mgNSBH KNe more often involve BH primaries of $\sim 3-4\,M_\odot$ rather than the $\sim 4\,M_\odot$ BH required for GW230529 to disrupt the NS and produce a KN. In addition, the O5 prior on the NS spin is broad, and the pairing allows the KN-producing mgNSBH population to extend to $m_2\simeq 2\,M_\odot$ even for the fiducial EOS. Starting from isolated binary formation channels, \cite{Drozda_2022} similarly finds that most NSBH KNe arise from mgNSBHs with relatively massive NSs. For more asymmetric NSBH systems, spin should exert an even stronger influence on tidal disruption than in GW230529 \cite{matur2024}; however, such asymmetric configurations are rare in our simulations because the pairing function suppresses extreme mass ratios (e.g., the rate for a BH five times the NS mass, as in \cite{matur2024}, is suppressed by a factor $\sim 4$ relative to more equal–mass systems).

\subsubsection{Viewing angle and distance distributions}

The viewing–angle distribution for both the mgNSBH population and GW230529 follows the Schutz distribution \cite{Schutz_2011} due to the GW selection effects, peaking near $\sim 30^\circ$. For GW230529, the viewing angle is effectively prior–dominated and thus unconstrained. The luminosity distance for O5 mgNSBH detections peaks around $\sim 530$\,Mpc for the fiducial EOS.

\subsubsection{Ejecta mass predictions}

For the O5 mgNSBH KN population, as it is clear from Figure \ref{fig:ejecta}, the dynamical ejecta mass cover a similar range to that expected for GW230529, with most support in the range $10^{-3}-10^{-2}~M_\odot$. Wind ejecta are typically more massive for the O5 population, peaking around $10^{-2}~M_\odot$ as more massive remnant masses are allowed.

\begin{figure*}[htpb!]
    \centering
    \includegraphics[scale=0.45]{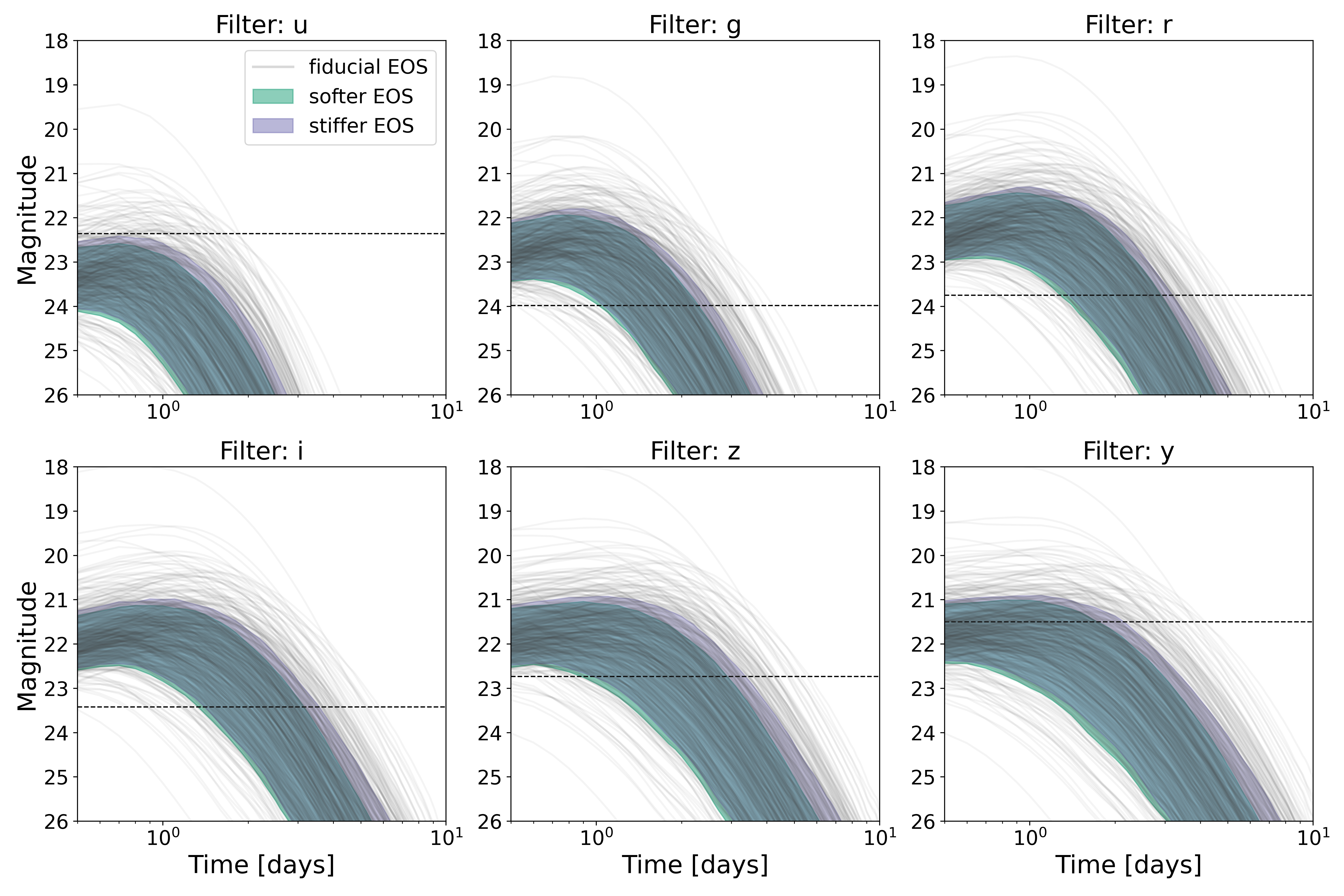}
    \includegraphics[scale=0.45]{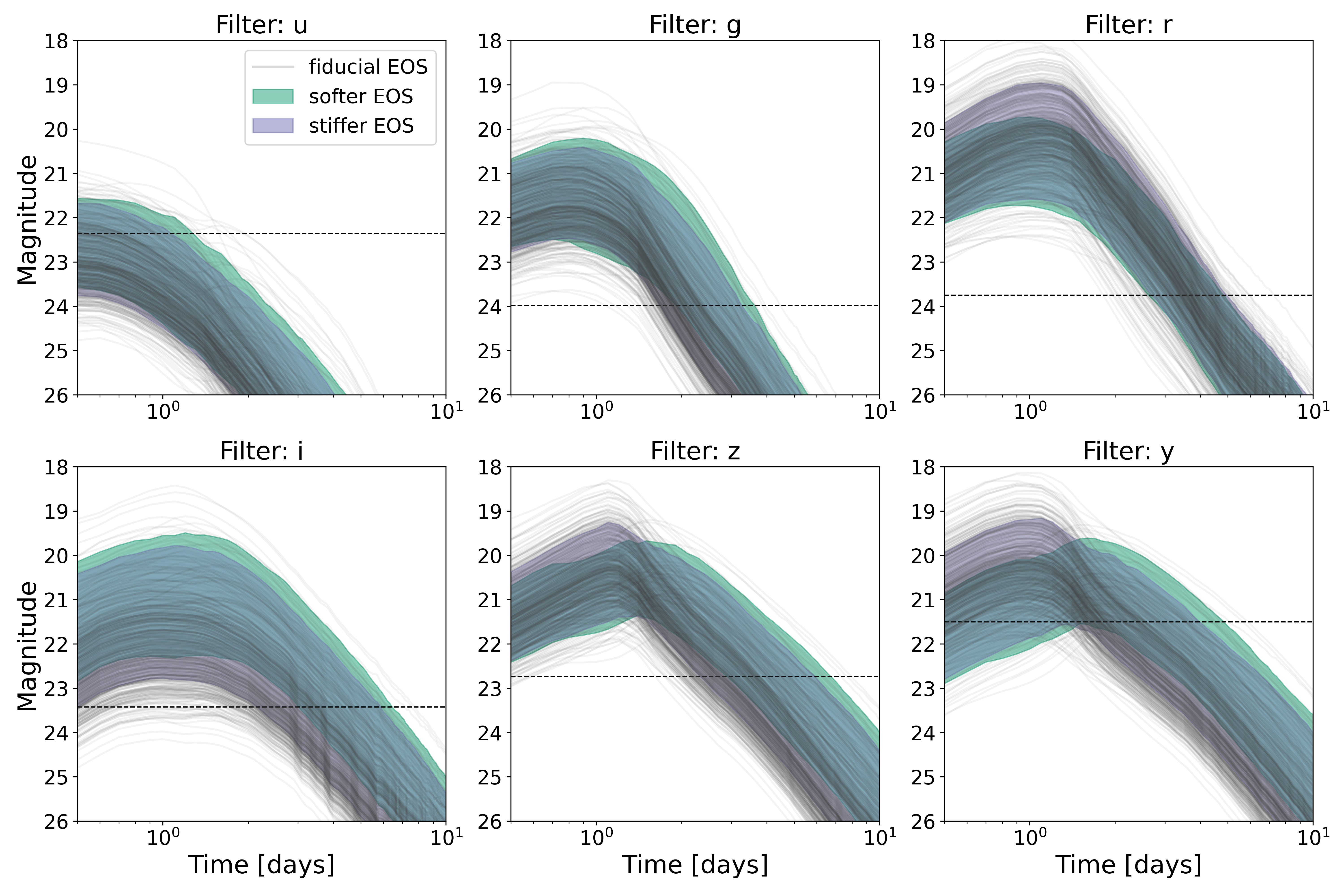}
        \caption{Simulated KN light curves from 0.5–10\,days post–merger in $ugrizy$ for GW230529 assuming NSBHs–BBHw (top six panels) and BNSs–BNSw (bottom six panels). Black curves show 500 randomly sampled KNe lightcurves for the fiducial EOS; shaded regions (green: softer; violet: stiffer) span the 10th–90th percentiles. The black dotted line marks a representative DECam $5\sigma$ depth for 90\,s exposures in each filter.}
    \label{fig:lc_bbh_bns}
\end{figure*}

\subsection{Kilonova detection}\label{sec:kndetection}

We now connect the population-level inferences to light curve predictions. We generate $ugrizy$ light curves up to 10\,days post–merger for the three EOS choices. We begin by examining the day–1 magnitude distributions for the three GW230529 posterior models, for the LVK O5 mgNSBH simulations, and for the LVK O5 BNS simulations (Figure~\ref{fig:hist_day1}), each sampled at distances $0$-$600$\,Mpc. %To enable a fair comparison with GW230529 reported at a median distance of $\sim 200$\,Mpc \cite{230529_LVK} we restrict the O5 simulated populations in this figure to $150$–$250$\,Mpc. 
Figure~\ref{fig:hist_day1} compares the distribution of $g$ and $i$ band magnitudes at 1\,day post–merger for GW230529 (under the NSBHs–BBHw, NSBHs–NSBHw, and BNSs–BNSw assumptions) against the O5 mgNSBH and BNS populations at $0-600$\,Mpc, for softer, fiducial, and stiffer EOSs. The Dark Energy Camera (DECam; \cite{Flaugher_2015}) reference depth provides a practical benchmark for detectability.

In both $g$ and $i$, the expected O5 BNS KN distributions show a tail that is as bright as the GW230529 BNSs–BNSw distribution. The GW230529 BNSw posterior favors relatively high masses compared to the generic O5 BNS population at the same distance and therefore tends to yield less ejecta for a given EOS, shifting its KN to fainter magnitudes. The O5 BNS distance distribution is flatter and extends to larger distances than GW230529, producing a fainter tail at high apparent magnitudes. For the softer EOS, the small subset ($\sim 1\%$) of GW230529 BNS posterior samples that produce a KN exhibits bright day–1 magnitudes, driven by the large wind–ejecta masses discussed in the previous subsections. Overall, the O5 BNS population typically produces brighter KNe than the mgNSBH population at comparable distances.

\begin{table*}%[!htpb]
    \centering
    \begin{tabular}{c|c| c | c | c | c | c |c}
        \hline
        \hline
        Event/ & Source type & \multicolumn{6}{c}{KN Percentage} \\
        population & & \multicolumn{6}{c}{detectable by DECam} \\
        \hline
        & & $u$ & $g$ & $r$ & $i$ & $z$ & $y$ \\
        \hline
        \hline
        \multirow{2}{*}{GW230529} & BNSs-BNSw &  21  &  100  &  100  &  97  &  100  & 98 \\
            & NSBHs-BBHw & 14 & 99 & 99 & 99 & 98 & 49 \\ 
            & NSBHs-NSBHw & 18 & 99 & 99 & 99 & 98 & 51 \\
        \hline
        LVK O5 & mgNSBH & 3 & 55 & 65 & 69 & 47 & 6\\ 
        \hline        
        \hline
    \end{tabular}
    \caption{Probability that the GW230529 KN could have been detected with DECam at 5$\sigma$ depth limit for a 90s exposure time for fiducial EOS, assuming different posterior models. We also show probabilities for a generic mass gap population from our LVK O5 simulation.}
    \label{tab:DECam detections expected}
\end{table*}

For GW230529 under the NSBH assumption, the NSBHs–BBHw and NSBHs–NSBHw models yield very similar KN light curves. The median day–1 peak magnitudes for NSBHs–NSBHw are $\sim 22.8$, $22.6$, and $22.5$ in $g$ and $\sim 21.8$, $21.7$, and $21.6$ in $i$ for the softer, fiducial, and stiffer EOS, respectively. The $i$–band is typically brighter than $g$ for NSBH KNe. The broader O5 mgNSBH population also favors KNe with $g\!\lesssim\!26$ and $i\!\lesssim\!25$, implying observability by large aperture telescopes.

Figure~\ref{fig:lc_bbh_bns} presents full $ugrizy$ light curves for GW230529 from $0.5$ to $10$\,days post–merger. We start at 0.5 days post-merger since the surrogate models work best in the $0.5-10$ days time range. NSBH KNe are typically red, peak between days 1 and 2, and reach their highest brightness in $z$. With the DECam $5\sigma$ reference depth for 90\,s exposures (dotted lines), we expect broad detectability in all bands except $u$. DECam, a wide–field, high performance CCD imager on the 4\,m Blanco telescope has been central to O4 GW follow–up efforts by GW–MMADS (Gravitational Wave Multi-Messenger Astronomy DECam Survey; \cite{cabrera2024,Hu2025}). Table~\ref{tab:DECam detections expected} presents DECam (90,s, gray time) detection probabilities for (i) a GW230529-like kilonova across different posterior assumptions and (ii) the O5 mgNSBH population, both evaluated under the fiducial EOS. For both BNSs–BNSw and NSBHs–BBHw, the $u$–band yields the lowest detection fraction, while the other bands typically reach $\sim 97$–$100\%$ (BNSs–BNSw) or $\sim 49$–$99\%$ (NSBHs–BBHw). A shallower search to $m\!\sim\!20.5$, such as the ZTF search of \cite{Ahumada_2024}, would have been unlikely to detect the predicted KN.

For the O5 mgNSBH simulations, we note that the KN detection efficiency is lowest in $u$ and highest in $i$. Up to $69\%$ of mgNSBH KNe should be detectable with DECam–like instruments in $i$; efficiencies in other bands range from the $6-65\%$ (Table~\ref{tab:DECam detections expected}).

EOS-dependent differences in NSBH light curves are modest: the stiffer EOS is up to $\sim 0.3$\,mag brighter at peak in $g$ than the softer EOS, reflecting more efficient NS disruption. The magnitudes evolve slowest in $zy$ and fastest in $u$, with the peak typically occuring near day~2 in $rzy$, day~1 in $u$ and $g$, and around $1.5$\,days in $i$. Because NSBHs–BBHw and NSBHs–NSBHw produce very similar light curves, we show only the BBHw case in Figure~\ref{fig:lc_bbh_bns}.

By contrast, the GW230529 BNSs–BNSw case exhibits greater diversity. The BNS light curves often peak by $\sim 8$\,hours post–merger and evolve more rapidly at blue wavelengths. Even so, assuming sufficient ejecta, the BNS scenario still offers a significant chance of detection with DECam–like instruments, up to 100\% probability of detection. Notably, the predicted light curves differ by orders of magnitude between the BNS and NSBH interpretations; thus, a KN detection coincident with a GW230529–like event can discriminate between these progenitor classes when GW data alone remain ambiguous \cite{Barbieri_2019}. In general, NSBH KNe may outshine those from high–mass BNSs \cite{Barbieri_2019,Kawaguchi_2020}, since high–mass BNS mergers tend to produce smaller dynamical ejecta and disks.

\subsection{Comparison with other works}

The GW230529 discovery article \cite{230529_LVK} computes the expected remnant mass outside the final BH, which we also compute here to estimate the NSBH ejecta masses, and can be used as a probe of KN production. After marginalizing over a range of EOSs, \cite{230529_LVK} report a $10\%$ probability that NS material is disrupted outside the BH for the NSBHs–BBHw posteriors. This aligns with our inferred $2$–$20\%$ KN–production fraction when we adopt the same posteriors. For the low–spin prior BBHw posteriors (which we did not analyze in this work), \cite{230529_LVK} find a $4\%$ disruption probability.

In Ref.~\cite{Chandra:2024ila}, the authors assess plausible formation mechanisms for GW230529 using available prior models, determine which component (NS or BH) formed first, and characterize the remnant BH from posterior samples. Our focus differs: we target KN emission and the regions of parameter space that yield detectable EM counterparts. For an expected mgNSBH population from isolated binaries, they compute ejecta masses using four EOSs: APR4 \cite{PhysRevC.58.1804}, SLy \cite{Chabanat:1997un}, DD2 \cite{PhysRevC.81.015803}, and H4 \cite{PhysRevD.79.124032}, whereas we generate KN light curves using the maximum–posterior EOS from \cite{Huth:2021bsp} and its upper/lower $95\%$ credible bounds to quantify the impact of stiffer/softer EOS on KN detectability for both GW230529 and a mgNSBH population. They predict peak KN apparent magnitudes for the mgNSBH population spanning $\sim 26.6-24.0$ from $u$ to $y$ bands, typically dimmer than our predictions for the EOS set we consider. We attribute the discrepancy to different population assumptions and KN/ejecta models, while noting that our O5 mgNSBH magnitude distributions are broad and overlap with their findings (see Figure~\ref{fig:lc_bbh_bns}).

In \cite{zhu2024}, the authors analyze GW230529 with the combined posterior \texttt{Combined\_PHM\_lowSecondarySpin} and compare it against GW200105 and GW200115 to probe formation channels and possible EM counterparts of GW230529. Our analysis also uses \texttt{Combined\_PHM\_lowSecondarySpin}, but we augment it with two additional posteriors that assume NSBH and BNS waveforms in order to capture the range of potential KN outcomes from this merger. They infer a high KN–production probability (up to $60\%$) and predict KNe up to $\sim 1$\,mag dimmer than our results (though less discrepant than the difference with \cite{Chandra:2024ila}). We attribute these differences to distinct EOS choices, ejecta–mass fitting formulae, and KN modeling frameworks. For the expected mgNSBH population, \cite{zhu2024} adopt \texttt{COMPAS}, a rapid binary population synthesis code \cite{Compas2022}. In this work, we take a data–driven approach: we assume compact–object populations that follow the PDB model \cite{Farah_2022} calibrated to LVK detections.

\section{Conclusion}\label{sec:conclusion}

In this article, we investigated the possible KN emission from the GW event GW230529, analyzing BBH, NSBH, and BNS waveform posteriors to capture different possibilities regarding the binary’s nature. For both the BBH and NSBH waveform posteriors, we assume NSBH KN production. In that case, we infer a $2$–$28\%$ probability of producing a KN, depending on the EOS. If instead GW230529 was a BNS, the KN emission probability lies between $0$ and $10\%$. We explicitly include the $0\%$ lower bound for KN production even though our calculations yield a minimum of $1\%$ to account for low spin prior BNS posteriors whose primary masses exceed the maximum NS mass for the EOSs considered and therefore cannot produce a KN.

We explored where KN-producing posterior samples reside in the parameter space. KN production favors lower $m_2$, but the $m_1$-$m_2$ degeneracy in the GW inference associates low $m_2$ with higher $m_1$, which in turn disfavors tidal disruption. As a result, across all posterior models, the KN-producing subset clusters around low $m_2$ and high $m_1$ ($\sim 1.4$ and close to $4~M_\odot$ respectively). The NSBHs–NSBHw models yield slightly larger probability of KN production than NSBHs–BBHw for two reasons: (i) the BBHw posterior permits lower (more negative) BH spins, which enlarge the ISCO and suppress disruption; and (ii) the NSBHw posterior provides greater support at $m_{2}\!\le\!1.4\,M_\odot$, and lower $m_2$ promotes KN production.

To contextualize GW230529 within a broader population, we simulated expected mass–gap NSBH (mgNSBH) mergers for the LVK O5 run using the \textsc{Power Law + Dip + Break} model and compared their properties to GW230529. The $\chi_1$ distribution in the O5 mgNSBH population shows a slight preference for positive alignment. Without the chirp–mass–driven $m_1$-$m_2$ degeneracy, KN production in the O5 mgNSBH population tends to be supported at lower $m_1$ ($\sim 3$–$4\,M_\odot$) than in the GW230529 NSBH scenarios (which favor $\sim 4\,M_\odot$).
  
For the GW230529 BNSs–BNSw interpretation, $6\%$ of samples produce a KN with our fiducial EOS. Of those, $\sim 97$–$100\%$ would be detectable with DECam in $griz$ using 90\,s exposures, with day–1 peaks at $g,i<23$. If GW230529 produced an NSBH KN and occurred in an observable region, there is a $\sim 98$–$99\%$ chance it would have been detectable with a DECam–like instrument in $griz$, with peak magnitudes around $g<24$ and $i<23$. Typical peak times are $\sim 1$\,day post–merger in $u,g$, $\sim 1.5$\,days in $r,i$, and $\sim 2$\,days in $z,y$.

Relative to GW230529, KN emission from O5 mgNSBH mergers is typically dimmer in the optical bands and can extend out to $\sim 600$\,Mpc. For ground–based follow–up, it would be wise to trigger follow-up observations at the redder wavelengths: $i$ deliver the highest detection probabilities ($\sim 69\%$), followed by $r$ and $g$ at $\sim 65$ and $55\%$, evaluated over the $\sim 3\%$ of mgNSBH mergers that produce a KN for the fiducial EOS. Given an expected O5 mgNSBH detection rate of $\sim 63$ per year under our distribution \cite{Farah_2022} and selection criteria, we anticipate $1-2$ mgNSBH KN per year with a substantial chance of detection. Taken together, these results indicate that mgNSBH mergers are poised to become a promising class of multimessenger sources in the near future.

\acknowledgments
\noindent \input{ack}

\nopagebreak
\appendix
\section{Ejecta mass fitting formulae}\label{sec:appendix}

The equations used to calculate the ejecta masses are summarized below. For BNS mergers, we follow the relation from \cite{Kr_ger_2020} to model the dynamical ejecta mass $M_{\rm dyn}$:
\begin{equation}
\label{eq:mdyn_bns}
    \frac{M_{\rm dyn}^{\rm BNS}}{10^{-3}} = \left (\frac{a}{C_1} + b \left (\frac{M_2}{M_1} \right)^{n} +cC_{1}\right)M_{1} + (1 \leftrightarrow 2).
\end{equation}
Here, the compactness $C_1$ is $GM_{1}/R_{1}c^{2}$, where $M_1$ and $R_1$ are the mass and radius of the primary (similarly for the secondary), and the best-fit parameters from the numerical simulations are $a= -9.3335$, $b = 114.17$, $c = -337.56$ and $n = 1.5465$ \cite{Kr_ger_2020}. 

The disk mass for BNS mergers follows the relation from \cite{Dietrich_2020}:
\begin{equation}
    \label{eq:mdisk_bns}
    \log_{\rm 10}\left(M_{\rm disk}^{\rm BNS}\right) = \max \left (-3, a\left(1 + b \tanh \left [\frac{c-{M_{\rm tot}}/{M_{\rm th}}}{d} \right] \right) \right), 
\end{equation}

where $M_{\rm th} = k_{\rm th}M_{\rm TOV}$ is the threshold mass \cite{Hotokezaka_2011},  $k_{\rm th}$ is a function of $M_{\rm TOV}$ and EOS \cite{Bauswein:2013jpa}, $a$ and $b$ are functions of the mass ratio $q$. These parameters are given by:
\begin{equation}
  \begin{split}
    &a =a_{0} + \delta_{a}x_{i} \\
    &b =  b_{0} + \delta_{b}x_{i}\\
    &x_i = 0.5 \tanh(\beta(q - q_{t})) \\
  \end{split}
\end{equation}
where $a_{0} = -1.581$, $b_{0} = -0.538$, $c = 0.953$, $d = 0.0417$, $\delta_{a}=-2.439$, $\delta_{b} = -0.406$, $\beta=3.910$, and $q_{t}=0.900$.

For NSBH mergers, the dynamical ejecta mass is given by the equation from \cite{Kr_ger_2020}:
\begin{equation}
    \label{eq:M_dyn_nsbh}
    \frac{M_{\rm dyn}^{\rm NSBH}}{M_{\rm NS}^{b}} = a_{1}Q^{n_{1}}\frac{1-2C_{\rm NS}}{C_{\rm NS}} - a_{2}Q^{n_{2}}\frac{R_{\rm ISCO}}{M_{\rm BH}} + a_{4}
\end{equation}
where the best fitting parameters are $a_1 = 0.007116$, $a_2 = 0.001436$, $a_4 =-0.02762$, $n_1 = 0.8636$ and $n_2 = 1.6840$. Here $Q={M_{\rm BH}}/{M_{\rm NS}}$ where the mass ratio is defined as $q=1/Q$ and $R_{\rm ISCO}$ is the radius of the innermost stable circular orbit (ISCO) of the black hole with mass $M_{\rm BH}$ and spin $\chi_{\rm 1z}$ in the direction of the orbital angular momentum. The baryonic mass of the NS is given by $M_{\rm NS}^{b}$ = $M_{ \rm NS} \left(1 + \frac{0.6C_{\rm NS}}{1-0.5C_{\rm NS}}\right)$ \cite{Lattimer_2001}, where $C_{\rm NS}$ is the compactness of the NS.

For NSBH mergers, we compute the remnant (baryon) mass remaining outside the BH at
$\sim 10$\,\emph{ms} post–merger using the fitting formula of \cite{Foucart_2018}:\cite{Foucart_2018}:
\begin{equation}
    \label{eq:mdisk_nsbh}
    \hat{M}^{\rm NSBH}_{\rm rem} = \left [{\rm max}\left( a\frac{1-2C_{\rm NS}}{\eta^{1/3}} - b~R_{\rm ISCO}\frac{C_{\rm NS}}{\eta} + c ,0\right) \right]^{1+d}.
\end{equation}

Here $\eta$ = $Q/(1+Q^{2})$ is the symmetric mass ratio, $\hat{M}$ = $M^{\rm NSBH}_{\rm rem}/M^{\rm b}_{\rm NS}$ and $a=0.40642158$, $b=0.13885773$, $c=0.25512517$, $d=0.761250847$.We obtain the NSBH disk mass by subtracting the dynamical component from the total remnant mass,
\begin{equation}
M_{\rm disk}^{\rm NSBH}
= M_{\rm rem}^{\rm NSBH} - M_{\rm dyn}^{\rm NSBH}.
\end{equation}

\bibliography{references1}
\bibliographystyle{apsrev4-1}

\end{document}

%% file: ack.tex
\noindent 
We thank the NMMA team for useful discussions and for the development of the NMMA code framework. In particular, we want to thank Thibeau Wouters for guidance with training the new surrogate model. We also want to thank Andrew Toivonen for useful feedback on the manuscript.

AP acknowledges support for this work by NSF Grant No. 2308193. AMF is supported by the National Science Foundation Graduate Research Fellowship Program under Grant No. DGE-1746045.
The authors thank Thibeau Wouters for help with the kilonova models. This research used resources of the National Energy Research
Scientific Computing Center, a DOE Office of Science User Facility supported by the Office of Science of the U.S. Department of Energy
under Contract No. DE-AC02-05CH11231 using NERSC award
HEP-ERCAP0029208 and HEP-ERCAP0022871. This work used resources on the Vera Cluster at the Pittsburgh Supercomputing Center. 
TD acknowledges funding from the Daimler and Benz Foundation for the project “NUMANJI” and from the European Union (ERC, SMArt, 101076369). Views and opinions expressed are those of the authors only and do not necessarily reflect those of the European Union or the European Research Council. Neither the European Union nor the granting authority can be held responsible for them. 

This research has made use of data or software obtained from the Gravitational Wave Open Science Center (gwosc.org), a service of the LIGO Scientific Collaboration, the Virgo Collaboration, and KAGRA. This material is based upon work supported by NSF's LIGO Laboratory which is a major facility fully funded by the National Science Foundation, as well as the Science and Technology Facilities Council (STFC) of the United Kingdom, the Max-Planck-Society (MPS), and the State of Niedersachsen/Germany for support of the construction of Advanced LIGO and construction and operation of the GEO600 detector. Additional support for Advanced LIGO was provided by the Australian Research Council. Virgo is funded, through the European Gravitational Observatory (EGO), by the French Centre National de Recherche Scientifique (CNRS), the Italian Istituto Nazionale di Fisica Nucleare (INFN) and the Dutch Nikhef, with contributions by institutions from Belgium, Germany, Greece, Hungary, Ireland, Japan, Monaco, Poland, Portugal, Spain. KAGRA is supported by Ministry of Education, Culture, Sports, Science and Technology (MEXT), Japan Society for the Promotion of Science (JSPS) in Japan; National Research Foundation (NRF) and Ministry of Science and ICT (MSIT) in Korea; Academia Sinica (AS) and National Science and Technology Council (NSTC) in Taiwan.
The authors are grateful for computational resources provided by the LIGO Laboratory and supported by National Science Foundation Grants PHY-0757058 and PHY-0823459.